\documentclass[12pt,epsf,amssymb]{article}

\usepackage{graphicx}
\usepackage{amsfonts}
\usepackage{amsmath}
\usepackage{amssymb}
\usepackage{amsthm}
\usepackage[mathscr]{eucal} 
\usepackage{url}

\usepackage[nosort]{cite}

\def\Z{\mathbb{Z}}
\def\Q{\mathbb{Q}}
\def\R{\mathbb{R}}
\def\C{\mathbb{C}}

\newcommand{\nin}{\mbox{\ooalign{\hfil/\hfil\crcr$\in$}}}

\newcommand{\vev}[1]{ \left\langle {#1} \right\rangle }

\newcommand{\GEV}{ {\rm GeV} }
\newcommand{\TEV}{ {\rm TeV} }

\def\tr{\mathop{\rm tr}}

\theoremstyle{definition}
\newtheorem{thm}{Theorem}[subsection]
\newtheorem{defn}[thm]{Definition}
\newtheorem{exmpl}[thm]{Example}
\newtheorem{props}[thm]{Proposition}
\newtheorem{lemma}[thm]{Lemma}
\newtheorem{anythng}[thm]{}

\newcommand{\p}[1]{\left(#1\right)}
\renewcommand{\b}[1]{\left\{#1\right\}}
\newcommand{\abs}[1]{\left|#1\right|}

\def\Omegabar{{\overline{\Omega}}}
\def\Sigmabar{{\overline{\Sigma}}}

\def\taubar{{\overline{\tau}}}

\def\Qbar{{\overline{\Q}}}

\newcommand{\dfn}[1]{\textbf{#1}}

\newcommand{\cM}{\mathcal{M}}

\newcommand{\cO}{\mathcal{O}}

\theoremstyle{remark}
\newtheorem*{prf}{Proof}
\hypersetup{
    colorlinks=true,
    linkcolor=black,
    citecolor=black,
    filecolor=black,
    urlcolor=black,
}

\begin{document}

\baselineskip 0.6cm

\begin{titlepage}
\begin{flushright}
 IPMU17-0080
\end{flushright}

\vskip 1cm

\begin{center}
  
  {\large \bf Revisiting arithmetic solutions to the $W=0$ condition}
  
  \vskip 1.2cm
 
Keita Kanno and Taizan Watari
 
 \vskip 0.4cm
  
{   Kavli Institute for the Physics and Mathematics of the Universe, 
   the University of Tokyo, Kashiwa-no-ha 5-1-5, 277-8583, Japan
}

 \vskip 1.5cm
   
 \abstract{The gravitino mass is expected not to be much smaller than the Planck scale for a large fraction of vacua in flux compactifications. There is no continuous parameter to tune even by hand, and it seems that the gravitino mass can be small only as a result of accidental cancellation among period integrals weighted by integer-valued flux quanta. DeWolfe et.al. (2005) proposed to pay close attention to vacua where the Hodge decomposition is possible within a number field, so that the precise cancellation takes place as a result of algebra. We focus on a subclass of those vacua---those with complex multiplications---and explore more on the idea in this article. It turns out, in Type IIB compactifications, that those vacua admit non-trivial supersymmetric flux configurations if and only if the reflex field of the Weil intermediate Jacobian is isomorphic to the quadratic imaginary field generated by the axidilaton vacuum expectation value. We also found that flux statistics is highly enriched on such vacua, as F-term conditions become linearly dependent. } 
 \end{center}
 \end{titlepage}
 
% \tableofcontents

%%%%%%%%%%%%%%%%%%%%%%%%%%%%%%%%%%%%%%%%%%%%%%%%%%%%%%
\section{Introduction}
%%%%%%%%%%%%%%%%%%%%%%%%%%%%%%%%%%%%%%%%%%%%%%%%%%%%%%

It has been known for more than a decade \cite{Denef:2008wq} that it is difficult 
to achieve both i) small cosmological constant and ii) natural supersymmetric 
unification under iii) the current understanding of string compactification. 
The supersymmetric unification scenarios prefer that the Kaluza--Klein scale of 
the internal geometry is not comparable to the string scale, but slightly 
lower \cite{Witten:1985xc}, and then in such a geometric phase, there is no 
good reason for the vacuum expectation value (vev) of the superpotential $\vev{W}$ 
to be much smaller than ${\cal O}(1)$ in Planck units.
That would immediately imply large gravitino mass, and also large supersymmetry 
breaking either in the F-term or D-term for the cosmological constant to be 
small.\footnote{When the K\"{a}hler potential and superpotential of the 4D 
effective theory is in the no-scale scenario, this argument for the F-term/D-term 
supersymmetry breaking is not applied. Also the anomaly mediated contributions 
to gauginos cancel, when the K\"{a}hler potential is in the sequestered form 
in addition (e.g., \cite{Bagger:1999rd}), so that the gravitino mass can be 
much larger than the gaugino mass. When the deviations from those assumptions 
are small, then the phenomenological requirement on the vev of 
$\int_X G \wedge \Omega$ is relaxed by that amount.} 

To be more specific, consider Type IIB string theory compactified on a Calabi--Yau 
threefold $X$ with Ramond--Ramond (RR) and NS--NS three-form fluxes $F^{(3)}$ and 
$H^{(3)}$. The superpotential and K\"{a}hler potential of the effective theory 
in 3+1-dimensions are given by \cite{DenefW}
\begin{align}
 W = \frac{M_{\rm Pl}^3}{\sqrt{4\pi}} \int_X G \wedge \Omega, \quad 
 \frac{K}{M_{\rm Pl}^2} \simeq
   - 2 \ln \left(\frac{{\rm vol.}(X)}{g_s^{3/2} \ell_s^6} \right)
   - \ln \left( i \int_X \Omega \wedge \overline{\Omega} \right)
   - \ln( -i(\phi - \phi^*)),
 \label{eq:eff-KW}
\end{align}
where $\phi$ is the dilaton chiral multiplet, ${\rm Im}(\vev{\phi}) = g_s^{-1}$, 
$G := F^{(3)} - \phi H^{(3)}$, and the three-form $\Omega(z)$ on $X$ depends on 
the complex structure moduli chiral multiplets denoted collectively by $z$. 
Everything in $\int_X G \wedge \Omega$ has been made dimensionless in the 
convention adopted in (\ref{eq:eff-KW}); once an integral basis 
$\{ e_{i=1,\cdots, b_3(X)} \}$ of $H^3(X;\Z)$ is chosen, then the fluxes are parametrized 
by integers (flux quanta) $\{n^{i=1,\cdots, b_3}\}$ and $\{ m^{i=1,\cdots, b_3} \}$, where
\begin{align}
G=\sum_{i=1}^{b_3} (n^i-\phi m^i) e_i;
\end{align}
period integrals 
\begin{align}
\Pi_i(z) = \int_{\gamma_i} \Omega(z)
\end{align}
are evaluated over the three-cycles $\gamma_i$ that are Poincar\'e dual to $e_i$; 
we therefore have a dimensionless combination 
$\int_X G \wedge \Omega(z) = \sum_i (n^i-\phi m^i)\Pi_i$. The gravitino mass 
$m_{3/2}$ is then given by 
\begin{align}
  \frac{m_{3/2}}{M_{\rm Pl}} = e^{\frac{K}{2M_{\rm Pl}^2}} \frac{W}{M_{\rm Pl}^3}
   = \pi^{1/4} \left( \frac{M_{6}}{M_{\rm Pl}} \right)^{\frac{3}{2}}
     \frac{ \int_X \langle G \wedge \Omega \rangle }
          { \sqrt{ \int_X i \vev{ \Omega \wedge \overline{\Omega}} } }\;(g_s)^{1/2}, 
\end{align}
where $M_{\rm Pl} \simeq 2.4 \times 10^{18} \; \GEV$ is the reduced Planck scale, and 
$M_6 := 1/R_6 := (\vev{{\rm vol}(X)})^{-1/6}$. When we think of supersymmetric 
unification scenarios where the Standard Model gauge groups originate from 7-branes 
wrapped on a 4-cycle with volume $\vev{{\rm vol}({\rm D7})}=: R_G^4$, the three 
observable parameters\footnote{
$M_{\rm GUT}$ is the energy scale of gauge coupling unification, and $\alpha_{\rm GUT}$ 
the value of the unified gauge coupling constant. } 
$M_{\rm Pl}$, $M_{\rm GUT}$ and $\alpha_{\rm GUT}$ are expressed 
in terms of three microscopic parameters of compactifications, $(g_s \ell_s^4)$, 
$R_6$ and $R_G$, which means that the factor $M_6/M_{\rm Pl}$ can be expressed
purely in terms of observable parameters (e.g., \cite{Tatar:2009jk}).
\begin{align}
  \pi^{1/4} \left( \frac{M_6}{M_{\rm Pl}} \right)^{\frac{3}{2}} \simeq 
    \frac{8.3 \times 10^{-4}}{c^2} 
    \left(\frac{M_{\rm GUT}}{2\times 10^{16} \; \GEV}\right)^2
    \left(\frac{1/24}{\alpha_{\rm GUT}}\right)^{\frac{1}{2}};
\end{align}
$c$ is a geometry-dependent factor of order unity in the relation $M_{\rm GUT} = c/R_G$.
Since the value of $g_s$ cannot be arbitrarily small even in Type IIB 
compactifications,\footnote{If the SU(5) symmetry breaking is due to
a line bundle in the geometric phase, 
$1 \lesssim (R_G/\ell_s)^4 = g_s/\alpha_{\rm GUT}$. The lower bound on $g_s$ 
is relaxed by $(2\pi)^{-4}$, if we require $1 \lesssim R_G/\sqrt{\alpha'}$.
The other geometric phase condition $1\lesssim (R_6/l_s)$, or 
$1\lesssim R_6/\sqrt{\alpha'}$, respectively, is slightly weaker. In F-theory, 
where the up-type quark Yukawa couplings are generated from $E_6$ 
algebra \cite{TWyukawa}, the value of $g_s$ is not an independent parameter.
}
the vev of $\int_X G \wedge \Omega$ needs to be small for the gravitino mass 
to be smaller than, say, $10^{-6} \times M_{\rm Pl}$.

Once a set of flux quanta $\{ n^i, m^i\}$ is specified, then the vev of 
$W \propto \int_X G \wedge \Omega$ is determined. Since the flux quanta can change 
only by an integer, it is not possible to tune the vev of $\int_X G \wedge \Omega$ 
continuously to a small value. At best we can hope that the $b_3$ contributions 
$\sum_{i=1}^{b_3} (n^i-\phi m^i) \Pi_i = \int_X G \wedge \Omega$ almost cancel 
one another accidentally. When $b_3$ complex numbers of order unity are generated 
randomly, and are summed up, the probability that the absolute value of the sum 
is $\epsilon$ or less is $\cO(\epsilon^2)$, which is small,\footnote{For
$m_{3/2}\sim10^3 \;\TEV$, $\p{g_s^{1/2} \cdot \epsilon} \sim 10^{-9}$ is necessary.}
but still non-zero and positive. In a large set of string flux vacua, therefore, 
there may still be a choice of flux quanta $\{n^i, m^i\}$ where 
$\vev{\int_X G \wedge \Omega}$ just happens to be small, and our local universe 
just happens to be described by such a flux configuration. Certainly such a story 
cannot be ruled out, but one can hardly say that it is a {\it natural} scenario 
of moduli stabilization behind supersymmetric unification. This is a problem 
that has been known for more than a decade \cite{Denef:2008wq}. While the absence 
of positive support in the LHC data at 13TeV does not kill low-energy supersymmetry 
scenarios immediately, this decade-old problem alone is enough to suggest that 
we are supposed to abandon either ii) natural supersymmetric unification or 
iii) the current understanding of flux compactifications of string theory.
If we are to keep ii), some drastic idea is necessary in modifying the current 
understanding of flux compactification.

Incidentally, there are some pioneering papers seeking for chances of arithmetics 
to play some roles in string theory (for examples, \cite{moore1998attractors, 
moore1998arithmetic,CdO1, CdO2, Schimmrigk, gukovvafarcft, moore2004houches, DeWolfe, 
DeWolfe2}). DeWolfe et.al. \cite{DeWolfe} in particular (see 
also \cite{moore2004houches}), introduced an idea that arithmetics of complex 
structure moduli vev plays some role in getting the vev\footnote{So far as 
the K\"{a}hler moduli are stabilized by non-perturbative effects, this 
$\int_X \vev{G \wedge \Omega}$ is by far the dominant contribution to 
$\vev{W}$. Although $\vev{W}$ does not have to be strictly zero 
phenomenologically, $\vev{W}$ will be precisely zero in the approximation 
of ignoring volume stabilization (non-perturbative effects) and supersymmetry 
breaking, if there is any {\it natural} solution to small $\vev{W}$.}  
$\vev{W} \propto \int_X \vev{G \wedge \Omega} =0$.  The idea is to focus on 
a subset ${\cal M}_{{\rm alg}}$ in ${\cal M}_{\rm cpx}^X \times {\cal M}_{\rm dil.}$, 
and think only of flux vacua in it. Here, ${\cal M}_{\rm cpx}^X$ is the moduli 
space of complex structure of a family of Calabi--Yau threefolds topologically 
the same as $X$, ${\cal M}_{\rm dil.} = {\rm SL}(2;\Z) \backslash {\cal H}_{g=1}$ 
that of the dilaton chiral multiplet. 
${\cal M}_{{\rm alg}}={\cal M}^X_{{\rm alg}}\times {\cal M}^{\rm dil.}_{{\rm alg}}$ is 
defined to be the set of points in ${\cal M}_{\rm cpx}^X \times {\cal M}_{\rm dil.}$ 
where all of $\vev{\phi} \in {\cal H}_{g=1}$, $\vev{\Pi_i}$ and $\vev{D_a \Pi_i}$ 
for $a=1,\cdots, h^{2,1}(X)$ and $i=1,\cdots, b_3(X)$ take values in some number 
field $K_{\rm tot}$. For a given set of vevs $\vev{z,\phi}$, the $\vev{W}=0$ 
condition can be regarded as an extra linear condition on flux quanta. If 
$\vev{z,\phi}\nin{\cal M}_{{\rm alg}}$, only the trivial flux configuration 
satisfies the $\vev{W}=0$ condition; for non-trivial fluxes, $\vev{W}\neq 0$ 
without an accidental fine cancellation as discussed earlier.
For an algebraic case $\vev{z,\phi}\in {\cal M}_{{\rm alg}}$, however,
the $\sum_{i=1}^{b_3} (n^i-\phi m^i) \Pi_i=0 \in K_{\rm tot}$ condition can be regarded 
as a {\it finite} number of $\Q$-coefficient linear conditions on $n^i$'s and $m^i$'s; 
this is because $K_{\rm tot}$ is $d_{K{\rm tot}}:=[K_{\rm tot}:\Q]$-dimensional vector 
space over $\Q$. When the space of fluxes satisfying the F-term condition 
at $\vev{z,\phi}$ forms a lattice $\Z^\kappa$, the $\vev{W}=0$ condition is therefore 
satisfied by fluxes in a sublattice $\Z^{\kappa-d_{K{\rm tot}}}$ \cite{DeWolfe}.
The F-term conditions can also be regarded as $d_{K{\rm tot}}\times (h^{2,1}+1)$ of 
$\Q$-coefficient linear conditions on the flux quanta in this case, and the lattice 
$\Z^\kappa$ has a rank \cite{DeWolfe}
\begin{align}
 {\kappa} = 2b_3 - d_{K{\rm tot}} \times (b_3/2).
  \label{eq:dW-formula-F}
\end{align}
The rank $\kappa_0$ of the $W=0$ sublattice is then given by 
\begin{align}
 \kappa_0 = \kappa - d_{K{\rm tot}} = 2b_3 - d_{K{\rm tot}} \times (b_3/2) - d_{K{\rm tot}}.
   \label{eq:dW-formula-W}
\end{align}

There is no top-down justification for not thinking of string vacua outside ${\cal M}_{\rm alg}$.
While there is higher chance of finding a flux vacuum with $\vev{W}=0$ for 
$\vev{z,\phi} \in {\cal M}_{\rm alg}$, it appears to be a sheer fine tuning in the current 
understanding of string compactification to focus on ${\cal M}_{\rm alg}$ in the first place. 
A perspective behind the arithmetic idea above is to rely more on bottom-up clues i) and ii) 
at the beginning of introduction, and to believe that the current understanding of string 
compactification overlooks something; future development of string theory will justify focusing 
on ${\cal M}_{\rm alg}$ or even on its smaller subset. Whether such a strategy is fruitful is yet 
to be seen. For now, the authors of this article intend to elaborate more 
on the arithmetic idea for the $\vev{W}=0$ problem, and prepare for further developments  
in the future. 

In this article, we explore two questions associated with this idea. The first question is 
what the subset ${\cal M}_{\rm alg}^X$ is like in ${\cal M}_{\rm cpx}^X$, and 
what the number field $K_{\rm tot}^X$ is like there.\footnote{$K_{\rm tot}^X$ is the number field 
generated by the vev of $\Pi_i$'s and $D_a\Pi_i$'s over $\Q$; $K_{\rm tot} = K_{\rm tot}^X(\phi)$.}  
The second question is how the estimation (\ref{eq:dW-formula-F}, \ref{eq:dW-formula-W}) 
of the rank $\kappa$ and $\kappa_0$ of 
supersymmetric flux quanta is at work. It appears from (\ref{eq:dW-formula-F}, \ref{eq:dW-formula-W})
that there is virtually no chance for a non-trivial supersymmetric flux to exist ($\kappa_0 > 0$) 
when the number field $K_{\rm tot}$ is an extension over $\Q$ of degree-4 or higher;\footnote{Applying 
the same logic as in (\ref{eq:dW-formula-F}, \ref{eq:dW-formula-W}) to F-theory flux compactification, 
we would have $\kappa = [2(h^{3,1}+1)+h^{2,2}_H]-d_{K{\rm tot}} \times h^{3,1}$ and 
$\kappa_0 = \kappa - d_{K{\rm tot}}$. Since $h^{2,2}_H$ scales as $4 h^{3,1} + {\rm const}.$ 
for large $h^{3,1}$ \cite{Sethi:1996es, Klemmh22,TWtateyoko},
cases with $d_{K{\rm tot}} = [K_{\rm tot}: \Q] \leq 6$ do not have to be ruled out 
in this line of reasoning. } in fact, the authors of \cite{DeWolfe} discovered that there are some 
cases hinting that truth is stranger than the naive estimation (\ref{eq:dW-formula-F}, 
\ref{eq:dW-formula-W}). We focus on a subset ${\cal M}_{\rm CM}$ of ${\cal M}_{\rm alg}$, 
where plenty of math literature is available, and address the two questions above. 
% What appeared to be a miracle in \cite{DeWolfe} turns out to be a consequence of complex 
% multiplications on a Calabi--Yau.  

This article is organized as follows. Sections \ref{sec:math-bkgd}--\ref{ssec:BV-review} 
provide an update of what is being understood about CM points ${\cal M}_{\rm CM}$, which is 
a subset of ${\cal M}_{\rm alg}$ that has particularly good properties from the perspective 
of arithmetic geometry, so that we have a better feeling of what the subset 
${\cal M}_{\rm CM}$ of ${\cal M}_{\rm cpx}$ is like. 
In sections \ref{ssec:BV-just-F} and \ref{ssec:BV-also-W}, we pick up a family of (possibly) 
infinitely many CM points in the form of Borcea--Voisin Calabi--Yau threefolds, and 
study conditions of existence of non-trivial supersymmetric flux configurations. 
The analysis is also generalized for CM-type Calabi--Yau threefolds that are not Borcea--Voisin 
type in section \ref{sec:non-BV}. The formula for $\kappa$ and $\kappa_0$ for the CM points 
turn out to be quite different from the estimate (\ref{eq:dW-formula-F}, \ref{eq:dW-formula-W}) 
for generic points in ${\cal M}_{\rm alg}$; conditions for CM points to have $\kappa >0$ or 
$\kappa_0 >0$ are stated in terms of reflex fields, which is also a concept important in 
arithmetic geometry. Discussions in section \ref{sec:discussions} include important observations 
that are not contained in the earlier sections. 

In this preprint, the appendices \ref{sec:field} and \ref{sec:Hodge} provide a quick overview of mathematical 
facts in field theory, Hodge theory and complex multiplications, on which the main text is based. 
We did not try to make this article fully self-contained, however. Such materials as class field 
theory, Shimura variety, Mumford--Tate group, field of definition and more, 
are necessary in appreciating the result of our study in sections \ref{sec:BV}--\ref{sec:discussions},
but not so much in following up the analysis; for this reason, we just provide references 
for those materials, and decided not to provide explanations 
or sometimes even definitions of jargons in this article. 
\\

\noindent {\bf Note added:} The Phys. Rev. D version of this article has adopted a style of presentation
that will be more friendly to physicists. Following the suggestion from the referee, we also have inserted
a brief discussion on orientifold projection as section 3.3 in the journal version. On the other hand,
some of the materials in this preprint version, the entire section 2.2 and part of the appendix B.1,
have been dropped from the journal version;
they are mostly concerned about the first question mentioned above, which involves too heavy
mathematics for Phys. Rev. D.

%%%%%%%%%%%%%%%%%%%%%%%%%%%%%%%%%%%%%%%%%%%%%%%%%%%%%%%%%%%
\section{Math Background}
\label{sec:math-bkgd}
%%%%%%%%%%%%%%%%%%%%%%%%%%%%%%%%%%%%%%%%%%%%%%%%%%%%%%%%%%%

%%%%%%%%%%%%%%%%%%%%%%%%%%%%%%%%%%%%%%%%%%%%%%%%%%%%%%
\subsection{CM Points and Arithmetics}
%%%%%%%%%%%%%%%%%%%%%%%%%%%%%%%%%%%%%%%%%%%%%%%%%%%%%%%

Let us first focus on the complex structure moduli space ${\cal M}_{\rm cpx}^X$ of a family of 
Calabi--Yau threefolds; a member of this family---a Calabi--Yau threefold---is denoted by $X$. 
Whenever we refer to ``a Calabi--Yau threefold'' in this article, it is implied that a specific 
complex structure (a point in ${\cal M}_{\rm cpx}^X$) is chosen already. The complex structure of 
a Calabi--Yau threefold $X$ introduces a Hodge decomposition on the vector space $H^3(X;\Q)$ over $\Q$. 
$H^{3,0}(X;\C)$ is the one-dimensional subspace of $H^3(X;\C)$ generated by $\Omega$, and 
$H^{0,3}(X;\C)$ by $\overline{\Omega}$; 
let $\{ \Sigma_{a=1,\cdots, h^{21}} \} \subset H^3(X;\Q) \otimes \C$ be a basis of $H^{2,1}(X;\C)$.
Overall, 
\begin{align}
  \left\{ \Omega, \Sigma_{a=1,\cdots, h^{21}}, \overline{\Sigma}_{a=1,\cdots, h^{21}}, \overline{\Omega} \right\}
\end{align}
forms a $\C$-basis of the vector space $H^3(X;\Q) \otimes \C$. 
Here, $\dim_\C ({\cal M}_{\rm cpx}^X) = h^{2,1}(X)$, and $b_3 = 2(h^{2,1}(X) + 1)$.
When we say that all the complex numbers $D_a \Pi_i$ with $a=1,\cdots, h^{2,1}$ and $i=1,\cdots, b_3$ 
take values in a number field $K_{\rm tot}$, we mean by this condition that there exists 
a $\C$-basis $\{ \Sigma_{a=1,\cdots, h^{2,1}} \}$ of $H^{2,1}$ so that 
$\vev{\gamma_i, \Sigma_a } =: \Sigma_{ai} \in K_{\rm tot}$. 

It is a highly non-trivial condition that a number field $K_{\rm tot}$ exists so that 
all the complex numbers $\Pi_i$, $\Sigma_{ai}$, $\overline{\Sigma}_{ai}$ and $\overline{\Pi}_i$ are 
contained within $K_{\rm tot}$. In the large complex structure moduli region, for example, 
the components of $\Omega$ are in the form 
\begin{align}
 \left( \Pi_i \right)^T & \; = \left( 1, t^a, \partial_{t^a} {\cal F}, 2{\cal F} - t^c \partial_{t^c} {\cal F} \right)^T, \\
 &  {\cal F} = \frac{\kappa_{abc}}{6} t^at^bt^c
   - \frac{n_{ab}}{2} t^a t^b - \frac{c_b}{24} t^b 
   - \frac{\zeta(3)}{(2\pi i )^3}\frac{\chi(X^\circ)}{2} 
   + \sum_{\beta \geq 0} \frac{n_\beta}{(2\pi i)^3} {\rm Li}_3(e^{2\pi i \vev{t, \beta}}), 
\end{align}
where $t^a$'s ($a=1,\dots,h^{2,1}$) are a set of local coordinates on ${\cal M}_{\rm cpx}^X$
and $\kappa_{abc}$, $n_{ab}$, $c_a$, $\chi(X^\circ)/2$ are known to be integers, and 
$n_\beta \in \Q$.
It is hopeless to try to argue whether a complex number given by a series expansion is 
algebraic or transcendental. When we require that $\Sigma_{ai}$'s are also algebraic, 
we have little idea how to find out systematically where in ${\cal M}_{\rm cpx}^X$ the conditions 
are satisfied. 

The Gepner point\footnote{Gepner points in moduli spaces of Calabi--Yau 
compactifications are specific points both in the K\"{a}hler and complex structure 
moduli spaces. Because we only refer to complex structure of Calabi--Yau's 
in this article, we use an expression ``a Gepner point in ${\cal M}_{\rm cpx}^X$'' 
for what is actually be ``a point in ${\cal M}_{\rm cpx}^X$ on which a Gepner model is 
projected to.'' } in the $h^{2,1}=101$-dimensional family of quintic Calabi--Yau 
threefolds is known to be in ${\cal M}_{\rm alg}^X$ (e.g., \cite{borceacm, shioda}), 
and so is the Gepner point in the $h^{2,1}=1$-dimensional family of the mirror quintics.
The corresponding number fields $K_{\rm tot}^X$ are cyclotomic fields. Complex 
structures of other Gepner models are also expected to have this property 
(cf. \cite{yui2012}). Given the enormous variety in the topology of Calabi--Yau 
threefolds, however, collection of one point from ${\cal M}_{\rm cpx}^X$ from some 
choices of topology appears to be a very small subset of all the possible complex 
structures in Type IIB compactifications.
Another class of examples is to construct $X$ through an orbifold of a product of 
three elliptic curves $E_{i=1,2,3}$, each one of which has a period $\tau_i$ that is 
algebraic \cite{DeWolfe}. How far can we go beyond this list?

Math literatures available at this moment do not seem to say much about 
${\cal M}_{\rm alg}^X$. There is, however, a subset of ${\cal M}_{\rm alg}^X$ that has 
long attracted interest of mathematicians for its significance in number theory. 
It is the set of CM points in ${\cal M}_{\rm cpx}^X$, which we review shortly, and 
there is plenty of math literatures available, so that there is a better hope 
to understand this subset---denoted by ${\cal M}_{\rm CM}^X$---systematically. It is 
also expected in string theory that all the rational conformal field theories 
with $(c, \tilde{c})=(9,9)$ central charges and ${\cal N} = (2,2)$ supersymmetry 
on world-sheet correspond to the points in ${\cal M}_{\rm CM}^X$ \cite{gukovvafarcft}.
While the bottom-up idea for the $W=0$ problem does not motivate to take the 
K\"{a}hler moduli vev also to be of CM-type, it would not sound too stupid to 
expect that better understanding on string theory in the future indicates that 
we must use rational CFT's for a consistent string compactifications; this 
would not only justify to choose the complex structure vevs to be of CM-type, but 
also predict that the K\"{a}hler moduli vevs also are.  

An elliptic curve of CM-type is an elliptic curve $E_\tau$ whose complex structure 
parameter $\tau$ in the upper complex half plane ${\cal H}_{g=1}$ is a solution 
to a non-trivial $\Q$-coefficient quadratic polynomial equation 
\begin{align}
\label{eq:quadratic-eq-tau}
 a\tau^2+b\tau + c=0, \qquad a,b,c \in \Z.
\end{align}
The collection of such $\tau$'s in ${\cal H}_{g=1}$, denoted by 
${\cal M}_{\rm CM}^{\rm ell.}$, is a very small subset of 
${\cal M}_{\rm alg}^{\rm ell} = {\cal H}_{g=1} \cap \overline{\Q}$ because 
$\tau \in {\cal M}_{\rm alg}^{\rm ell.}$ can be a solution to a $\Q$-coefficient 
polynomial of any degree, not necessarily quadratic. 
The subset ${\cal M}_{\rm CM}^{\rm ell.}$ of ${\cal M}_{\rm alg}^{\rm ell.}$ has a very nice 
property, however; $j(\tau)$ is algebraic, and hence one can choose all the 
coefficients of the defining equation of $E_\tau$ to be algebraic numbers 
(e.g., III.1.4, \cite{silvermanAEC}): 
\begin{align}
 y^2 + xy = x^3 -\frac{36}{j(\tau)-1728} x - \frac{1}{j(\tau)-1728}.
\end{align}
This means that $E_\tau$ has a model over a number field $K=\Q(\tau, j(\tau))$. 
The converse is also known to be true (Thm. IIc, \cite{schneider1937arithmetische}): 
when $E_\tau$ is defined over a number field, and $\tau \in {\cal M}_{\rm alg}^{\rm ell.}$, 
then $\tau \in {\cal M}_{\rm CM}^{\rm ell.}$. Elliptic curves with complex multiplication 
can therefore be characterized as those where both $\tau$ and $j(\tau)$ are algebraic.

This observation is not limited to the case of elliptic curves. The definition of 
CM type on complex structure has been generalized from elliptic curves to Abelian varieties 
of higher dimensions, K3 surfaces and Calabi--Yau threefolds (see the appendix \ref{ss:cmperiod} 
for a brief review, or references therein for more). At least it is now known, when $X$ 
is either an Abelian variety or a K3 surface in ${\cal M}_{{\rm alg}}$, that $X$ with a complex structure of CM-type is 
defined over a number field \cite{shimura1961complex, shimura2016abelian, pjateckii1975arithmetic, 
rizov2005complex}, and the converse is also true \cite{cohen1996humbert, shiga1995criteria, 
tretkoff2015k3}.\footnote{For some Calabi--Yau threefolds, see \cite{tretkoff2013transcendence, 
tretkoff2015transcendence}.}  

When a $d$-dimensional variety $X$ is defined over an algebraic number field, the Hasse--Weil 
$L$-function is defined in association with its $H^d(X)$, and even the zeta function 
$\zeta(X,s)$ is for $X$. When the Hodge structure on $H^d(X)$ is of CM-type, furthermore, 
it is expected\footnote{See (II.10, \cite{silvermanAdv}) for the case of elliptic curves, 
(\S 19, \cite{shimura2016abelian}) for Abelian varieties,  
(\S 1, \cite{pjateckii1975arithmetic}) for K3 surfaces, and \cite{yui2013modularity} for 
higher dimensions. See also \cite{CdO1, CdO2}.}
that the $L$-function factorizes into a product of $L$-functions each one of which 
is associated with an embedding of the CM endomorphism field of $H^d(X)$ into $\C$.
While it is not obvious whether the zeta function $\zeta(X,s)$ and its factorization property 
have a role to play in string theory, it will not be too surprising even if they are relevant to 
consistency of flux compactification of string theory. 

%%%%%%%%%%%%%%%%%%%%%%%%%%%%%%%%%%%%%%%%%%%%%%%%%%%%%%%
\subsection{Andr\'e--Oort Conjecture and Coleman--Oort Conjecture}
%%%%%%%%%%%%%%%%%%%%%%%%%%%%%%%%%%%%%%%%%%%%%%%%%%%%%%%

There are infinitely many CM points in the moduli space ${\cal M}_{\rm cpx}^{\rm ell.} = 
{\rm SL}(2;\Z) \backslash {\cal H}_{g=1}$ of elliptic curves. Any CM field $K$ with 
$[K:\Q] = 2$ (i.e., a quadratic imaginary field) appears as the endomorphism field 
at infinitely many CM points. There are also infinitely many CM points for a given 
CM type $(F,\Phi_F)$ with $[F:\Q] = 2g$ in the moduli space 
${\cal M}_{\rm cpx}^{A_g} = {\rm Sp}(2g;\Z) \backslash {\cal H}_g$ of Abelian varieties 
with complex $g$-dimensions. An enormous literature exists on this subject, 
because of its relevance to the class field theory \cite{morelandclass, milne1997class, ShimuraIntro,
bilucomplex}. There are also infinitely many CM-type K3 surfaces, with infinitely many 
variations for a given CM field $K$ so far as $[K:\Q] \leq 20$; much less seems to be known, 
though, about the list of CM fields available in the period domain $D(T_0)$ of a given even lattice 
$T_0$ with signature $(2,2n-2)$ (cf. the appendix \ref{ss:cmperiod} and references therein).

Those CM points arising in an infinite series are packed into\footnote{\label{fn:math-sources}The 
authors found \cite{milne2004introduction, kerr2010shimura} and \cite{zarhin1983hodge} 
useful in learning Shimura variety, connected Shimura variety, and Mumford--Tate group, 
respectively.} a Shimura variety ${\rm Sh}(MT, \tilde{h})$ associated with its Mumford--Tate 
group $MT$; a homomorphism $\tilde{h}$ in (\ref{eq:htilde}) just needs to be chosen from one 
of those infinitely many CM points. In the case of Abelian varieties, it is given by\footnote{
For an extension field $E$ over $F$, ${\rm Res}_{E/F}(\mathbb{G}_m)$ is an Abelian group 
identical to $E^\times = E \backslash \{ 0\}$. With the notation ${\rm Res}_{E/F}(\mathbb{G}_m)$, 
however, one emphasizes that this group is being regarded as an algebraic group defined over $F$.} 
(e.g., Rmk. 3.5, \cite{kerr2010shimura})
\begin{align}
MT = N_{\Phi^r}({\rm Res}_{K^r/\Q}(\mathbb{G}_m)) = N_{\Phi^r}((K^r)^\times),
  \label{eq:MT-reflex-rltn}
\end{align}
whereas $MT = K^\times = {\rm Res}_{K/\Q}(\mathbb{G}_m)$ in the case of K3 surfaces. 
The Mumford--Tate group is implemented as an algebraic subgroup over $\Q$ of 
$\mathbb{G}{\rm Sp}(2g)$ [resp. $\mathbb{G}O(2,2n-2)$] in the case of Abelian varieties 
[resp. K3 surfaces],\footnote{The group $\mathbb{G}{\rm Sp}(2g)$ [resp. $\mathbb{G}O(2,2n-2)$] 
is the set of linear transformations that preserve the skew symmetric bilinear form on $H^1(A;\Q)$ 
associated with the polarization of $A$ [resp. the intersection form on the lattice $T_0$] up to 
scalar multiplication. This fudge factor of scalar multiplication is introduced so that the image 
of $\tilde{h}$ in (\ref{eq:htilde}) can be accommodated. Using representations of 
$\C^\times \cong {\rm Res}_{\C/\R}(\mathbb{G}_m)$ instead of those of pure complex phase 
$S^1 \subset \C^\times$, descriptions in the language of algebraic geometry are made possible.} 
and such an implementation $(K^r)^\times \rightarrow MT \subset \mathbb{G}{\rm Sp}(2g)$ 
[resp. $(K)^\times \cong MT \subset \mathbb{G}O(2,2n-2)$] maps the Shimura variety of CM points 
into the moduli space of Abelian varieties [resp. K3 surfaces] through 
\begin{align}
{\rm Sh}(MT, \tilde{h}) \hookrightarrow {\rm Sh}(\mathbb{G}{\rm Sp}(2g),{\cal H}_g) & \rightarrow 
{\rm Sp}(2g;\Z) \backslash {\cal H}_g = {\cal M}_{\rm cpx}^{A_g}, \\
{\rm Sh}(MT,\tilde{h}) \hookrightarrow {\rm Sh}(\mathbb{G}O(2,2n-2),D(T_0)) & \rightarrow 
{\rm Isom}(T_0)\backslash D(T_0) = {\cal M}_{\rm cpx}^{{\rm K3}(T_0)}.
\end{align}
See \cite{milne2004introduction}.\footnote{\label{fn:math-sources2}We found that \cite{milne1997class, kerr2010shimura,clarkray}
are also useful in learning the relation between the class field 
theory and Shimura variety.} CM points in a connected Shimura variety (like ${\cal M}_{\rm cpx}^{A_g}$
and ${\cal M}_{\rm cpx}^{{\rm K3}(T_0)}$) show up in the form of a family, a zero-dimensional Shimura 
variety ${\rm Sh}(MT,\tilde{h})$ (as stated above). Andr\'e--Oort conjecture 
(cf \cite{moonenoort, tsimerman2015proof} and references therein) states, furthermore, that 
when a set of CM points ${\cal S}$ is in a connected Shimura variety ${\rm Sh}_K(G,X)$ 
associated with a Shimura datum $(G,X)$, then the Zariski closure of ${\cal S}$ is the union 
of finitely many connected Shimura subvarieties ${\rm Sh}_{K_i}(G_i, X_i)$.
If the set of CM points ${\cal S}$ contains representatives from infinitely many 
mutually distinct zero-dimensional Shimura varieties, in particular, 
some of the ${\rm Sh}_{K_i}(G_i, X_i)$'s should have a positive dimension. 

We should have a little different story for the complex structure moduli space ${\cal M}_{\rm cpx}^X$ 
of a family of Calabi--Yau threefolds, however. First, the space of polarized rational Hodge 
structure on $H^3(X;\Q)$ with the symplectic intersection form is a coset 
space (e.g., \cite{borceacm, kerr2015algebraic})
\begin{align}
D = {\rm Sp}(b_3;\R)^+ / {\rm U}(1) \times {\rm U}(h^{2,1}), 
\label{eq:CY3-period-domain}
\end{align}
where we understand that the Hodge--Riemann bilinear relations have been enforced. 
This space\footnote{\label{fn:careless-Gamma}
In this article, we do not pay much attention to the choice of a discrete group 
with which to take the quotient, and avoid technical details that are not 
essential to the central theme of this article.} is not regarded as a connected 
Shimura variety, but an example of a more general class of objects called 
Mumford--Tate domain in \cite{green2012mumford}. It is still expected 
(e.g., VIII.B, \cite{green2012mumford}), just like in the Andr\'e--Oort conjecture 
for a Shimura variety, that the Zariski closure of a set of CM points in a 
Mumford--Tate domain is the union of a finitely many Mumford--Tate subdomains, 
and hence appears as families of infinite many CM points \cite{borceacm}.

Secondly, the moduli space of a family of Calabi--Yau threefolds ${\cal M}_{\rm cpx}^X$ 
is an $h^{2,1}$-dimensional subvariety\footnote{See footnote \ref{fn:careless-Gamma}. 
We do not repeat this reminder in the rest of this article. } of $D$; note that 
$\dim_\C (D) = [(h^{2,1})^2 + 5h^{2,1}+2]/2  > h^{2,1}$. Even when a CM point with 
a given Mumford--Tate group $MT \subset \mathbb{G}{\rm Sp}(b_3)$ is found 
in ${\cal M}_{\rm cpx}^X \subset D$, other CM points in $D$ that share the same 
Mumford--Tate group are not necessarily in $\cM_{\rm cpx}^X$. This means that 
CM points do not necessarily arise as a family of infinitely many 
in ${\cal M}_{\rm cpx}^X$, although they do in $D$. The Andr\'e--Oort conjecture and 
its variation for Mumford--Tate domains still have a thing to say, however. 
If a set ${\cal S}$ of CM points in ${\cal M}_{\rm cpx}^X \subset D$ contains 
representatives from infinitely many mutually distinct zero-dimensional Mumford--Tate 
domains, then its Zariski closure should be the union of a finitely many Mumford--Tate 
domains, some of which must have a positive dimension. This implies, in particular, 
that ${\cal M}_{\rm cpx}^X$ needs to contain a Mumford--Tate subdomain of a positive 
dimension; in other words, the orbit of the group action in the Mumford--Tate 
subdomain must be aligned with ${\cal M}_{\rm cpx}^X$. Therefore, group theoretical 
considerations can be exploited in listing up possible Mumford--Tate subdomains 
contained in ${\cal M}_{\rm cpx}^X \subset D$.  

Now, which closed subvariety $Y$ of a Mumford--Tate domain $D$ does admit a 
Mumford--Tate subdomain? This question has been studied the best in the case of 
$D = {\cal H}_g$ and $Y={\cal T}_g$, the closure of the Torelli locus 
${\cal T}^\circ_g$, which is the subspace of ${\cal H}_g$ corresponding to the 
Jacobian of non-singular genus $g$ curves (e.g., see \cite{moonenoort, oortCMJac} 
for a review). This is a trivial question for $g \leq 3$, but is not for $4 \leq g$, 
since ${\cal T}_g$ is a proper subspace of ${\cal H}_g$ for those cases.
For the range of $4 \leq g \leq 7$, infinitely many CM points have been found 
in the Torelli locus ${\cal T}^\circ_g \subset {\rm Sp}(2g;\Z) \backslash {\cal H}_g$ 
\cite{dJnoot1991, shimura1964, oortIntro}, but it is expected (Coleman--Oort 
conjecture), that there will be at most finitely many CM points in 
${\cal T}^\circ_g \subset {\rm Sp}(2g;\Z)\backslash {\cal H}_g$ at least for large $g$ 
(maybe for $g \geq 8$) \cite{coleman1987conj, oort1997liftcm}; a proof is not known 
yet. Note that this expectation does not rule out infinitely many CM points present 
in ${\cal T}_g \backslash {\cal T}_g^\circ$. We will come back to this point 
in section \ref{sec:BV}.

%%%%%%%%%%%%%%%%%%%%%%%%%%%%%%%%%%%%%%%%%%%%%%%%%%%%%%
\section{Flux Vacua in a Borcea--Voisin Calabi--Yau Threefold}
\label{sec:BV}
%%%%%%%%%%%%%%%%%%%%%%%%%%%%%%%%%%%%%%%%%%%%%%%%%%%%%%%%

%%%%%%%%%%%%%%%%%%%%%%%%%%%%%%%%%%%%%%%%%%%%%%%%
\subsection{Borcea--Voisin Construction}
\label{ssec:BV-review}
%%%%%%%%%%%%%%%%%%%%%%%%%%%%%%%%%%%%%%%%%%%%%%%%

Borcea--Voisin Calabi--Yau threefolds were introduced \cite{borceacm} as families of 
Calabi--Yau threefolds where the alignment between ${\cal M}_{\rm cpx}^X$ and 
a Mumford--Tate subdomain of the period domain $D$ is more likely\footnote{In fact, 
all the examples examined in \cite{borceacm} are the $g=0$ cases (reviewed 
in the following), where the alignment does take place. 
See also \cite{garbagnati2010examples}.} than in general families of Calabi--Yau 
threefolds.\footnote{Another systematic construction of CM-type varieties (Viehweg--Zuo 
construction) has been reported \cite{VZ}; see also \cite{rohde2009cyclic, 
moonen2010cyclec} for additional information. The Fermat quintic Gepner point, 
for example, is not the only CM point in moduli space ${\cal M}_{\rm cpx}^X$ of the 
quintic threefolds, but this construction provides a set of CM points whose closure 
is a 2-dimensional subspace in the $(h^{2,1}=101)$-dimensional ${\cal M}_{\rm cpx}^X$. 
See \cite{moonenoort, oortCMJac} for variations.} Thus, there is a good chance in 
such families of finding numerous examples of CM-type Calabi--Yau threefolds, where 
the arithmetic solutions to the $\vev{W}=0$ condition can be implemented. 

Here is a brief summary of what is known about Borcea--Voisin Calabi--Yau 
threefolds \cite{borceacm, voisink3, borceak3}, mostly for the purpose of setting 
the notation to be used in the following. When $S$ and $E$ are a K3 surface and 
an elliptic curve, respectively, and there is an automorphism 
$\sigma_S: S \rightarrow S$ and $\sigma_E: E \rightarrow E$ so that 
$\sigma_S^*(\Omega_S) \wedge \sigma_E^*(\Omega_E) = \Omega_S \wedge \Omega_E$, we can 
think of an orbifold $(S \times E)/(\sigma_S, \sigma_E)$ while leaving unbroken 
supersymmetry in 3+1-dimensions. Since an automorphism $\sigma_E$ of an elliptic 
curve can only be of either order 2, 3, 4 or 6, the K3 surface $S$ needs to have 
a non-symplectic automorphism $\sigma_S$ where order is either 2, 3, 4 or 6. The 
original construction by Borcea and Voisin \cite{borceacm, voisink3, borceak3} was 
for cases with order 2 automorphisms $(\sigma_S, \sigma_E)$, and we also deal with 
those cases in this article; generalization of the following analysis must be 
straightforward.\footnote{
For F-theory applications, an orbifold of $S_1 \times S_2$ of a pair of K3 surfaces 
can be used instead. There is more freedom in the choice of automorphisms for the 
orbifold then. That would be more fruitful for particle physics applications, but 
in the present study, we restrict ourselves to Type IIB compactifications.} 

Nikulin carried out \cite{nikulin1983factor} classification\footnote{Similar 
classification of automorphisms of a K3 surface that are not order-2 has been 
studied in \cite{nikulin1983factor, vorontsov1983automorphisms, mukai1988finite, 
kondo1992automorphisms, oguiso1993remark, kondo1998niemeier, oguiso1998k3, 
oguiso1999k3, oguiso2000vorontsov, garbagnati2007symplectic, artebani2008non, 
taki2010non, schutt2010k3, artebani2011k3, taki2011classification, 
taki2012classification, taki2014oguiso, tabbaa2014classification, 
artebani2015symmetries} and references therein. } of lattice ``polarizations'' of 
a K3 surface that has an order-2 non-symplectic automorphism $\sigma_S$; 
to be more precise,\footnote{In order to preserve 
unbroken supersymmetry in 3+1-dimensions, we just need that the image of $\sigma_S \in {\rm Aut}(S)$ 
in the isometry group ${\rm Isom}(T)$ of the transcendental lattice $T$ has an order 
$m_T = 2$, $3$, $4$ or $6$. If $\sigma_S$ is an order $m$ element in ${\rm Aut}(S)$, then $m_T| m$, 
but there can be a non-trivial kernel---symplectic automorphisms---in the projection 
${\rm Aut}(S) \supset \Z/(m\Z) \rightarrow \Z/(m_T\Z) \subset {\rm Isom}(T)$. Examples of 
a K3 surface with such an automorphism group have also been found in the literatures 
in the previous footnote. }\raisebox{5pt}{,}\footnote{We should also keep in mind that 
a moduli $z_s$ may be in a Noether--Lefschetz locus (see footnote \ref{fn:NL}) 
within the period domain $D(T_0)$, where the transcendental lattice $T$ of a K3 surface 
may be a proper subset of the lattice $T_0$. For this reason, we maintain quotation marks 
in ``polarization'' here. See also footnote \ref{fn:non-NL-K3}.} let $NS_0 \oplus T_0$ 
be the pair of mutually orthogonal primitive lattices within $H^2(K3;\Z) \cong {\rm II}_{3,19}$ 
where $\sigma_S$ acts trivially on $NS_0$ and by $(-1) \times$ on $T_0$. There are 75 
such lattice ``polarizations''; detailed information of those lattice pairs are 
found in \cite{nikulin1983factor}.
Two integers $(r,a)$ extract important properties of those 75 lattice ``polarizations''.\footnote{
Here is a few notable examples of those 75 lattice pairs. The $(r,a)=(1,1)$ 
case corresponds to a degree-2 K3 surface, $(r,a)=(2,0)$ to an elliptic K3 surface, 
$(r,a)=(2,2)$ to $NS_0 = U[2]$, and their mirrors are $(r,a)=(19,1)$ with $T_0 = U \oplus \vev{+2}$, 
$(r,a)=(18,0)$ with $T_0 = U^{\oplus 2}$, $(r,a)=(18,2)$ with $T_0 = U \oplus U[2]$, respectively.
Those without a mirror include $(r,a)=(20,2)$ where $T_0 = \vev{+2} \oplus \vev{+2}$ and  
$(r,a)=(18,4)$ corresponding to a Kummer surface $S = {\rm Km}(E \times F)$.}
$r$ is the rank of $NS_0$, which means that ${\rm rank}(T_0)=22-r$. The Abelian group 
$NS_0^*/NS_0$ is isomorphic to $(\Z/2\Z)^a$ in this classification, and $a$ the number of generators 
of  $(\Z/2\Z)^a$
($NS_0^*$ is the dual lattice of $NS_0$).

The fixed locus of $E$ under the order-2 automorphism $\sigma_E = [(-1) \times]$ consists of 
four points. The set of points $S_{\sigma_S}$ of a K3 surface $S$ that are fixed
under an order-2 non-symplectic 
automorphism $\sigma_S$, on the other hand, consists of irreducible curves disjoint from one another. 
Apart from two cases\footnote{In one case, $S_{\sigma_S}$ is empty, while $S_{\sigma_S}$ consists 
of two elliptic curves in the other. We do not provide separate analysis of flux vacua for 
Borcea--Voisin Calabi--Yau's using one of the two cases of lattice polarization of K3 surface 
in this article.} out of the list of 75 lattice ``polarizations'',  
\begin{align}
% Z=
 S_{\sigma_S} = C_g + L_1 +\dots +L_k,
\end{align}
where $C_g$ is a curve with genus $g$ and $L_i$ are rational curves. It is known that the integers 
$k$ and $g$ are related to $r$ and $a$ through \cite{nikulin1983factor}
\begin{align}
k=\dfrac{1}{2}(r-a),\quad g=\dfrac{1}{2}(22-r-a).
\end{align}
The fixed point locus of an order-2 automorphism $(\sigma_S, \sigma_E)$ of $S \times E$ therefore 
consists of $Z := \cup_{b=1}^4 Z_b :=\cup_{b=1}^4 (C_{g(b)} + L_{1(b)} + \cdots + L_{k(b)})$,
where each $b$ corresponds to one of the four fixed points under $\sigma_E$ in $E$. 
The $(\sigma_S,\sigma_E)$-twisted sector fields of the orbifold $(S \times E)/(\sigma_S, \sigma_E)$ 
are localized along $Z$. 

By allowing\footnote{\label{fn:non-NL-K3}
With this, $T_0$ can be identified with the transcendental lattice $T_S$
and $NS_0$ with the N\'eron--Severi lattice $NS_S$ of a K3 surface $S$, if $S$ corresponds to a generic 
(non-Noether--Lefschetz)
point in this period domain. Although we might find motivations to maintain distinction 
between $NS_0$ and $NS_S$ in particle physics applications of F-theory compactifications, 
we remain simple minded in this article, and are happy to think just of a most generic point in $D(T_0)$.}
a K3 surface $S$ to take all possible complex structure in the period domain 
$D(T_0)$ of the lattice $T_0$, we find a $(20-r)$-dimensional space of complex structure for $S$. 
There is one more for $E$, $\tau \in {\cal H}_{g=1}$, as we just use an automorphism of order 2.
By turning on vacuum expectation values in the massless fields in the twisted sector, 
we would have $4g$ more parameters of complex structure deformation of this Calabi--Yau threefold; 
the complex structure deformation of the $\C^2/(\Z/(2\Z))$ singularity along $Z$ corresponds to 
global sections of $(1,0)$ form on the curve $Z$. This is how we obtain \cite{borceacm, voisink3, borceak3}
\begin{align}
h^{2,1}=1+(20-r)+4g.
\end{align}

Although one could think of a family of Calabi--Yau threefolds over an 
$h^{2,1}$-dimensional moduli space ${\cal M}_{\rm cpx}^X$, we would then have to face 
the alignment problem referred to at the end of the previous section. 
Instead, we focus on vacua that sit within a sub-family over the $(20-r)+1$-dimensional 
moduli space 
\begin{align}
 ({\rm Isom}(T_0) \backslash D(T_0)) \times ({\rm SL}(2;\Z) \backslash {\cal H}_{g=1}).
   \label{eq:moduli-BV}
\end{align}
That is to remain in the orbifold limit.\footnote{
We do not say anything about K\"{a}hler moduli in this article. 
So, the twisted sector fields corresponding to the resolution of 
the $\C^2/(\Z/(2\Z))$ singularity may well have non-zero vacuum expectation value. 
The K\"{a}hler moduli therefore have $h^{1,1}=1+r+4(k+1)$ degrees of freedom.}
This subspace is an example of connected Shimura varieties associated with 
the group $\mathbb{G}O(2,20-r) \times GL(2)$.

Let $X$ be the minimal resolution of the $\C^2/(\Z/(2\Z))$ singularity of the orbifold 
$(S \times E)/(\sigma_S, -1)$. The cohomology group $H^3(X; \Q)$ can be obtained as the 
$(\sigma_S, -1)$-invariant (even) part of $H^3({\rm Bl}_Z(S \times E);\Q)$ (Thm. 7.31, \cite{voisinhodge}). 
% (Th\'eor\`eme 7.31). 
First, 
\begin{align}
H^3({\rm Bl}_Z (S \times E) ;\Q) \simeq \left( H^2(S;\Q) \otimes H^1(E;\Q)\right) \oplus H^{1} (Z;\Z)
\end{align}
as a vector space over $\Q$, and secondly, its even part under $(\sigma_S, -1)$ is  
\begin{align}
 H^3(X;\Q) \cong (T_0 \otimes H^1(E;\Q)) \oplus \left( \oplus_{a=1}^4 H^1(C_{g(a)}; \Q) \right)
  =: V'_0 \oplus \left( \oplus_{a=1}^4 V_a \right),
\end{align}
where each $a$ component corresponds to one of the four twisted sectors.
In the case of a Borcea--Voisin Calabi--Yau threefold $X$ with $g \geq 1$, 
with its complex structure remaining in the orbifold limit, 
the rational Hodge structure on $H^3(X;\Q)$ has rational Hodge substructures on 
$V'_0$ and each one of $V_a$'s; if $g = 0$, then $V_a$'s are empty. We refer 
to $V'_0$ as the untwisted sector, and $V_a$'s the twisted sector.  
The rational Hodge substructure on $V'_0$ is level-3, and those of $V_a$'s level-1. 
The level-1 Hodge structure on $V_a$ is essentially the same as the weight-1 Hodge 
structure of a $g$-dimensional Abelian variety ${\rm Jac}(C_g)$. 

The Mumford--Tate domain (\ref{eq:CY3-period-domain}) contains a connected Shimura 
variety associated with the image of a homomorphism
\begin{align}
\mathbb{G}{\rm Sp}(b_3) \to \mathbb{G}O(2,20-r) \times GL(2) \times \left( \mathbb{G}{\rm Sp}(2g) \right)^4,
\end{align}
and the moduli subspace (\ref{eq:moduli-BV}) of ${\cal M}_{\rm cpx}^X$ should be 
mapped identically to the first two factors of a connected Shimura variety
\begin{align}
  D(T_0) \times {\cal H}_1 \times ({\cal H}_g)^4,  
\end{align}
while the map $D(T_0) \rightarrow {\cal T}_g \subset {\cal H}_g \subset ({\cal H}_g)^4$ for 
$g \geq 1$ is non-trivial. The condition that the Calabi--Yau threefold $X$ is of CM-type 
(see the appendix \ref{ss:cmperiod}) is equivalent to all the rational Hodge substructures 
on $V_0'$ and $V_a$'s being of CM-type. The rational Hodge substructure on $V'_0$ is of CM-type 
if and only if those on $T_0$ and $H^1(E;\Q)$ are of CM-type (Prop. 1.2, \cite{borceacm}). 
So, on a point 
\begin{align}
 (z_S, \tau) \in  {\cal M}_{\rm CM}^{{\rm K3}(T_0)} \times {\cal M}_{\rm CM}^{\rm ell.} \subset 
 D(T_0) \times {\cal H}_1  \subset {\cal M}_{\rm cpx}^X,
\end{align}
the Hodge structure on $V'_0$ is guaranteed to be of CM-type.
There remains a question whether a CM $z_S$ gives rise to an Abelian variety 
${\rm Jac}(C_g)$ with sufficiently many CMs or not. 
Therefore, the alignment problem between ${\cal M}_{\rm cpx}^X$ and a Mumford--Tate 
subdomain of $D$ in (\ref{eq:CY3-period-domain}) can be regarded as composition of 
the two following problems in the case of Borcea--Voisin Calabi--Yau threefolds: 
a) the map $D(T_0) \rightarrow {\cal T}_g \subset {\cal H}_g$, and 
b) the CM points within ${\cal T}_g$ (on which the Coleman--Oort conjecture has things to say 
for large $g$ cases). This alignment problem has been studied in math literatures, 
and at least it is known that a non-empty subset of CM points in $D(T_0)$ are mapped to 
CM points in ${\cal T}_g \subset {\cal H}_g$ for not a small number of lattice 
pairs (Cor. 3.5, \cite{yui2012}).\footnote{The study in \cite{yui2012} does 
not impose a condition that those non-empty subsets stay away from Noether--Lefschetz 
loci of $D(T_0)$.} 

In this article, we work out which rational Hodge structure admits non-trivial 
supersymmetric flux configurations in sections \ref{ssec:BV-just-F} 
and \ref{ssec:BV-also-W}, treating as if ${\cal H}_g$ (rational Hodge structure 
on the twisted sector $V_a$'s) is yet another moduli parameter independent 
of $D(T_0)$; this physics analysis needs to be combined with the math analysis of 
a) and b), in principle, but we will just make a brief remark on the result of 
our physics analysis and b) in this article. 

%%%%%%%%%%%%%%%%%%%%%%%%%%%%%%%%%%%%%%%%%%%%%%%%%%%
\subsection{Fluxes Satisfying the F-term Conditions}
\label{ssec:BV-just-F}
%%%%%%%%%%%%%%%%%%%%%%%%%%%%%%%%%%%%%%%%%%%%%%%%%%%%%%

There are $1+h^{2,1} = 1 + (20-r) + 1 + 4g$ chiral multiplets in the Type IIB compactification 
in question; one of them is the dilaton $\phi$. The complex structure moduli of $X$, denoted 
collectively by $z$ can be split into the twisted sector moduli $y_{ar}$ 
($a=1,\cdots, 4$, $r=1,\cdots, g$), $\tau$ for the elliptic curve $E_\tau$ and 
$(20-r)$ moduli for the K3 surface $S$ denoted collectively by $z_S$. As we seek for flux 
vacua that sit within the orbifold limit $D(T_0) \times {\cal H}_1 \subset {\cal M}_{\rm cpx}^X$, 
that is, $\vev{y_{ar}} = 0$, the F-term conditions with respect to $y_{ar}$'s are concerned about 
the Hodge structures on the twisted sector $V_a$'s, and those with respect to $\phi$, $\tau$ and 
$z_S$'s about the one on the untwisted sector $V'_0$. We can therefore work on the F-term 
conditions separately on $V_a$'s, and on $V'_0$. 

Let us first work on the F-term conditions on the twisted sector (which is irrelevant 
in the first place in the Borcea--Voisin Calabi--Yau's with $g=0$). 
Since the analysis remains precisely the same, word-by-word, for different isolated twisted 
sectors labeled by $a=1,\cdots, 4$, we will now drop the label $a$ for a while. 
Using a symplectic integral basis $\{f^{r=1,\cdots, g}, \underline{f}_{r=1,\cdots,g} \}$ of 
$H^1({\rm Jac}(C_g);\Q) \cong V_a$, flux quanta on $V_a$ are parametrized by rational numbers 
$\{n_r, \underline{n}^r \}$ and $\{ m_r, \underline{m}^r \}$ for the RR and NS--NS sectors, 
respectively. 
Let $\tau^{rs}(z_S) \in {\cal H}_g$ be the complex structure of the Abelian variety 
${\rm Jac}(C_g)$ so that the $(1,0)$ (or $(2,1)$) component of $V_a$ is spanned by 
$\{ (f^r + \tau^{rs} \underline{f}_s) \; | \; r=1,\cdots, g\}$. The F-term conditions 
for $y_{as}$ ($s=1,\cdots, g$) are written as 
\begin{align}
  \left( \underline{n}^r - \phi \underline{m}^r, -(n_r - \phi m_r) \right)
     \left( \begin{array}{c}  \delta_r^{\; s}\\ \tau^{rs} \end{array} \right) = 0 
   \qquad s = 1,\cdots, g.
\end{align}

We have assumed that the Abelian variety ${\rm Jac}(C_g)$ has sufficiently many 
complex multiplications. ${\rm Jac}(C_g)$ is further assumed to be 
isotypic (\ref{statmnt:def-isotypic}), for now, because we do not lose generality 
by doing so; we can just carry out the same analysis for individual isotypic components 
of $H^1({\rm Jac}(C_g);\Q)$. The algebra of Hodge endomorphisms of $V_a$ is then 
a CM field---denoted by $K^C$---with $[K^C:\Q] = 2g$. 
This means that the $g$ conditions above labeled by 
$s$ can be re-organized through a linear transformations in a number field 
$(K^C)^{\rm nc} \subset \overline{\Q}$ into $g$ conditions each one of which is associated 
with an embedding $\rho_a \in \overline{\Phi}$ of the CM type $(K^C, \Phi)$ of the Abelian 
variety ${\rm Jac}(C_g)$ (see \ref{statmnt:CMfield-repr-cohomology}). Moreover, 
the $g$ column vectors (labeled by $a=1,\cdots, g$) after the re-organization has the property
explained in the appendix \ref{ssec:reltn-Hdg-cmp-embd}, so there exists a $\Q$-basis 
$\{ y_{r=1,\cdots, g}, \underline{y}^{r=1,\cdots, g} \}$ of $K^C$ such that the F-term conditions are 
equivalent to  
\begin{align}
  \left( \underline{n}^r - \phi \underline{m}^r \right) \rho_a(y_r)
  - (n_r - \phi m_r) \rho_a(\underline{y}^r) = 0 \qquad a=1,\cdots, g.
\end{align}
These conditions can be further brought into 
\begin{align}
  \rho_a \left( \underline{n}^r y_r - n_r \underline{y}^r \right) = \phi 
  \rho_a \left( \underline{m}^r y_r - m_r \underline{y}^r \right),  \qquad 
  \rho_a \in \overline{\Phi}.
\end{align}
Here, $(\underline{n}^r y_r - n_r \underline{y}^r)$ and 
$(\underline{m}^r y_r - m_r \underline{y}^r)$ are regarded as elements in an abstract 
finite extension field $K^C$ over $\Q$, while the axi-dilaton vev $\phi$ is a complex number. 
Now here is a 

\begin{props}
\label{props:dilaton}
Let $A$ be an Abelian variety of $\dim_\C A = g$ with sufficiently many CMs, and 
$(K,\Phi)$ its CM type. When there is an element $x \in K$ such that 
$\rho_a(x) = \phi \in \C \backslash \R$ for all $\rho_a \in \overline{\Phi}$, then 
both $\phi$ and $x$ have a degree-2 minimal polynomial over $\Q$. In particular, 
$\Q(\phi)$ is a quadratic imaginary field.  
\end{props}

\begin{prf}
First, note that $x \nin K_0$, where $K_0$ is the totally real subfield of $K$, because 
$\rho_a(x)$'s do not fall into $\R$. This means that $K = K_0(x)$. Second, the 
fact that $K$ is a degree-2 extension over $K_0$ implies that there exist $P, Q \in K_0$ 
so that $x^2 = P x + Q$. The assumption that $\rho_a (x) = \phi \in \C$ for all 
$\rho_a \in \overline{\Phi}$ implies that 
\begin{align}
 \sum_{\rho_a \in \overline{\Phi}} \rho_a (x^2) & \; = \sum_{\rho_a \in \overline{\Phi}} \rho_a(P x + Q)
   = \sum_{\rho_a \in \overline{\Phi}} \left( \rho_a(P) \rho_a(x) + \rho_a(Q) \right), \\
 g \phi^2 & \; = {\rm Tr}_{K_0/\Q}(P) \phi + {\rm Tr}_{K_0/\Q}(Q),
\end{align}
where we used the fact that the complex conjugate pair of embeddings $\rho_a \in \overline{\Phi}$ 
and $\bar{\rho}_a \in \Phi$ become the same embedding upon restriction to $K_0$, and 
${\rm Tr}_{K_0/\Q}(a) = \sum_{\rho_a \in \overline{\Phi}}\rho_a(a) = 
\sum_{\bar{\rho}_a \in \Phi} \bar{\rho}_a(a)$ for $a \in K_0$.
As ${\rm Tr}_{F/\Q}(a)$ takes a value in $\Q$ for any $a \in F$ by definition, 
this shows that $\phi \in \C$ has a degree-2 minimal polynomial over $\Q$. 

Moreover, we see that $\rho_a(P) \in \R$'s are independent of $a=1,\cdots, g$ (and so are 
$\rho_a(Q) \in \R$'s), because $\phi^2 = \rho_a(x^2) = \rho_a(P) \phi + \rho_a(Q)$ holds 
in the quadratic imaginary field $\Q(\phi)$ for all $a = 1,\cdots, g$. In fact,
$\rho_a(P) = g^{-1} {\rm Tr}_{K_0/\Q}(P) \in \Q$, and $\rho_a(Q) = g^{-1} {\rm Tr}_{K_0/\Q}(Q) \in \Q$ 
independent of $a$, and the minimal polynomial of $x$ over $\Q$ is the same as that of $\phi$.
 $\bullet$
\end{prf}

This proposition is used in the analysis for the twisted sectors by setting $x = 
(\underline{n}^ry_r - n_r \underline{y}^r)/(\underline{m}^ry_r - m_r \underline{y}^r) \in K^C$. 
In doing so, we assume that the NS--NS flux is non-trivial\footnote{
\label{fn:stabilize} We wish to do so, 
in order to provide mass terms to the chiral multiplets $y_{ar}$'s.} in the twisted sector $V_a$, 
$a=1,\cdots, 4$, so $x$ is well-defined. For a non-trivial flux 
to be consistent with the F-term conditions in the twisted sector, therefore, the endomorphism 
field $K^C$ needs to satisfy 
\begin{align}
  K^C = K^C_0({}^\exists x_C), \qquad \Q(x_C) \cong \Q(\phi), \qquad [\Q(\phi):\Q] = 2.
  \label{eq:cond-F-1-BV}
\end{align}
Using the notion of the reflex field, the condition above can also be stated as 
\begin{align}
  (K^C)^r \cong \Q(\phi), \qquad [\Q(\phi) : \Q] = 2.
  \label{eq:cond-F-2-BV}
\end{align}

The reflex field of a quadratic imaginary field is the quadratic imaginary field itself, 
and we should also keep in mind that $(K^C)^{rr}$ is the CM field of the unique primitive 
CM subtype of the CM type $(K^C, \Phi_C)$ of ${\rm Jac}(C_g)$ (see the statement \ref{statmnt:Krr-Kprime}). 
This does not mean anything for $g=0$ and $g=1$ cases; for the cases with $g \geq 2$, however, 
the condition (\ref{eq:cond-F-2-BV}) with $(K^C)^{rr} = (K^C)^r$ implies that ${\rm Jac}(C_g)$ 
is not a simple Abelian variety, but isogenous to a product of $g$ elliptic curves each 
one of which is isogenous to $E_\phi$ (cf Prop. 27, \cite{shimura1961complex, shimura2016abelian}).
Such a Jacobian variety as ${\rm Jac}(C_g) = E_\phi \times \cdots \times E_\phi = (E_\phi)^g$ is not 
for a non-singular genus $g$ curve $C_g$, but for a degenerate limit of $C_g$ splitting into $g$ 
elliptic curves, and hence is not within the Torelli locus ${\cal T}_g^\circ$. Such a conclusion 
from physics analysis (conditions for presence of supersymmetric flux configurations) is not 
in conflict with the Coleman--Oort conjecture, but rather in line with it, even for large $g$ cases. 

Let us now work on the F-term conditions in the untwisted sector. For simplicity
(and for genericity), we assume that the rational Hodge structure on 
$V'_0 \cong T_0 \otimes H^1(E;\Q)$ is simple. The F-term conditions on this untwisted 
sector, from $\tau$, $z_S$'s and $\phi$, require that the $(1,2)$ and $(3,0)$ Hodge 
components of the flux $G = F^{(3)}-\phi \; H^{(3)}$ vanish. This is precisely equivalent to the analysis 
for the twisted sector, when the Abelian variety ${\rm Jac}(C_g)$ is replaced by the 
Weil intermediate Jacobian associated with the weight-3 Hodge structure on 
$V'_0 = T_0 \otimes H^1(E;\Q)$. Therefore, for a non-trivial flux configuration 
to exist on $V'_0$ consistently with the F-term conditions (cf footnote \ref{fn:stabilize}),
the following conditions need to be satisfied:\footnote{When two fields $K_1$ and $K_2$ are 
subfields of a common field $L$, the \dfn{composite field} $K_1 K_2$ is the minimal subfield 
of $L$ containing $K_1K_2$. 
This definition of $K_1K_2$ requires that we know a priori that such a field $L$ exists. 
We implicitly assume here that the rational Hodge structure on $V'_0$ is simple, which 
guarantees that $L$ exists. When the rational Hodge structure is not simple, 
as in section \ref{ssec:BV-also-W}, we need separate treatment.  
}
\begin{align}
  (K^SK^E)^r \cong \Q(\phi), \qquad [Q(\phi):\Q] =2.
  \label{eq:cond-F-3-BV}
\end{align}
When the twisted sector is non-empty (i.e., $g \geq 1$), the reflex fields of 
the untwisted and the twisted sectors need to be one and the same quadratic imaginary field 
$\Q(\phi)$ associated with the dilaton vev.

Suppose now that the endomorphism fields on a CM point satisfy the conditions stated above. 
First of all, the number field $K_{\rm tot} \subset \overline{\Q}$ in \cite{DeWolfe} in this case 
becomes 
\begin{align}
 K_{\rm tot} = K_{\rm tot}^X = (K^S)^{\rm nc} K^E (K^C)^{\rm nc}; 
\end{align}
$\Q(\phi)$ is included in $(K^S)^{\rm nc}$ and $(K^C)^{\rm nc}$.

Secondly, we are now ready to find the space of flux quanta consistent with the F-term 
conditions with this complex structure. We can choose arbitrary NS--NS flux quanta 
$\{ \underline{m}^{ar}, m_{ar} \}_{a=1,\cdots,4 ; r =1,\cdots, g}$ in the twisted sector without 
violating the F-term conditions; once the NS--NS flux quanta is specified, however, 
the RR flux quanta appropriate for getting the right $\phi$ vev is uniquely determined. 
Similarly, we can choose NS--NS flux quanta arbitrarily in the untwisted sector, 
though the RR flux configuration is now uniquely determined to get the right $\phi$ vev. 
Overall, 
\begin{align}
  \kappa = 2(22-r)+ 4 \times 2g = b_3(X).
  \label{eq:kappa-BV}
\end{align}
%

%%%%%%%%%%%%%%%%%%%%%%%%%%%%%%%%%%%%%%%%%%%
\subsection{Fluxes Satisfying both the F-term and $\vev{W}=0$ Conditions}
\label{ssec:BV-also-W}
%%%%%%%%%%%%%%%%%%%%%%%%%%%%%%%%%%%%%%%%%%%

Let us now impose the $\vev{W}=0$ condition. Since complex structure moduli and dilaton vev have been fixed 
by the F-term conditions for a given flux configuration, it is a yes-or-no question to ask whether $\vev{W}=0$ 
or not for a given flux configuration. The space of flux configuration satisfying $\vev{W}=0$ as well 
is a subspace of the $\kappa$-dimensional space of fluxes satisfying the F-term conditions for a given 
$(z, \phi) \in {\cal M}_{\rm cpx}^X \times {\cal M}_{\rm dil.}$.  

As is known well in the literature, the $\vev{W} \propto \int G \wedge \Omega =0$ condition and the F-term 
condition $D_\phi W = 0$ can be reorganized into 
\begin{align}
\int_X F^{(3)} \wedge \Omega =0 \qquad \text{ and } \qquad 
\int_X H^{(3)} \wedge \Omega =0.
\end{align}
The additional constraint $\vev{W}=0$ therefore implies that there is a linear relation 
among the algebraic numbers (e.g., \cite{moore2004houches,DeWolfe2,AspinwallFixing,TWFK3K3})
\begin{align}
 \sum_{i=1}^{2(22-r)} n^i \Pi_i = 0.
\label{eq:extra-linear-dep}
\end{align}
If the rational Hodge structure on $V'_0 = T_0 \otimes H^1(E;\Q)$ were simple, and the 
Hodge structure on $T_0$ and $H^1(E;\Q)$ were of CM-type, then the algebra of Hodge 
endomorphisms on $V'_0$ would be a CM field of degree $2(22-r) = \dim_\Q V'_0$, and 
the period integrals $\Pi_{i=1,\cdots, 2(22-r)}$ would be the image of the elements of 
a $\Q$-basis of the CM field under the embedding associated with the $(3,0)$ Hodge 
component. The existence of a non-trivial $\Q$-linear relation among the images of 
the embedding of a basis of the CM field is an outright contradiction. The extra 
condition $\vev{W}=0$ requires that the rational Hodge structure on 
$V'_0 = T_0 \otimes H^1(E;\Q)$ is not simple. 

In the case of Borcea--Voisin Calabi--Yau threefold $X$, the rational Hodge structure 
on $V'_0$ can be made not to be simple, when the CM point $(z_S, \tau)$ satisfies 
an extra condition (\ref{eq:cond-W0-1}). $V'_0$ is split into two vector subspaces 
over $\Q$, $V'_0 \cong V_0 \oplus \underline{V}_0$ with $\dim_\Q (V_0) = 
\dim_\Q (\underline{V}_0) = (22-r)$, and both $V_0$ and $\underline{V}_0$ have  
a rational Hodge substructure of $V'_0$. Let us verify this claim in the following. 

Let $\{ e'_1,\cdots, e'_{22-r} \}$ and $\{ \hat{\alpha}, \hat{\beta} \}$ be a basis of 
the vector space $T_0$ and $H^1(E;\Q)$ over $\Q$, respectively, and 
\begin{align}
\Omega_S=
\begin{pmatrix}
e'_1& \dots & e'_{22-r}
\end{pmatrix}
\begin{pmatrix}
\epsilon(y_1)\\
\vdots\\
\epsilon(y_{22-r})
\end{pmatrix}
,\quad 
\Omega_E=
\begin{pmatrix}
\hat{\alpha} & \hat{\beta} 
\end{pmatrix}
\begin{pmatrix}
1\\
\tau
\end{pmatrix},
  \label{eq:SandE-period-def-cmpnt}
\end{align}
where $\epsilon: K^S \hookrightarrow \C$ is the embedding associated with the $(2,0)$ Hodge 
component of $T_0 \otimes \C$, and $\{ y_1, \cdots, y_{22-r}\}$ is the basis of $K^S$ that is introduced 
in the appendix \ref{ssec:reltn-Hdg-cmp-embd}. The corresponding holomorphic three-form on $X$ 
is given by 
\begin{align}
\Omega = 
\begin{pmatrix}
e'_{i=1,\cdots, 22-r}\wedge \hat{\alpha} & e'_{i=1,\cdots, 22-r}\wedge \hat{\beta} 
\end{pmatrix}
\begin{pmatrix}
\epsilon(y_i) \\
\tau \cdot \epsilon(y_i)
\end{pmatrix}, 
 \label{eq:OmegaX-BVCY-cmpnt}
\end{align}
where differential forms $e'_i \wedge \hat{\alpha}$ and $e'_i \wedge\hat{\beta}$ on 
$S \times E$ should be pulled back to ${\rm Bl}_Z(S \times E)$, and then be regarded\footnote{
We only discuss in $\otimes \Q$ and do not pay attention to integrality in this article. So, 
this abuse of notations does not lead to any practical problem.} 
as those on $X$. The condition (\ref{eq:extra-linear-dep}) implies that there is a non-trivial 
set of rational numbers $\b{n'_i,n''_i}_{i=1,\dots ,22-r}$ so that 
\begin{align}
\begin{pmatrix}
n'_1 & \dots & n'_{(22-r)} & n''_1 & \dots & n''_{22-r}
\end{pmatrix}
\begin{pmatrix}
\epsilon(y_i) \\
\tau \cdot \epsilon(y_i)
\end{pmatrix}=0.
\end{align}
This means that there is an element 
\begin{align}
 \xi_S := \frac{n'_i y_i}{n''_i y_i} \in K^S
\end{align}
so that $\epsilon(\xi_S) = \tau$. As a first lesson, we see that 
\begin{align}
 K^E = \Q(\tau) \subset \epsilon(K^S) \subset (K^S)^{\rm nc} \qquad 
  \left( {\rm which~also~means~that~} K_{\rm tot} = (K^S)^{\rm nc}(K^C)^{\rm nc} \right)
\end{align}
in order for the extra condition $\vev{W}=0$ to be satisfied in a non-trivial
($n'' \neq 0$) flux vacuum.

To find out the decomposition of the rational Hodge structure on $V'_0$, as 
claimed earlier, it is useful to take a $\Q$-basis $\{ \alpha_{i=1,\cdots, (22-r)/2} \}$ of 
the totally real subfield $K^S_0$, and use $\{ \alpha_{i=1,\cdots, (22-r)/2}, \xi_S \alpha_{i=1,\cdots, (22-r)/2} \}$
as a $\Q$-basis of $K^S$, instead of $\{ y_{i=1,\cdots, (22-r)} \}$. Changing the basis of $T_0$ from 
$\{ e'_i \}$ to $\{ e''_i \}$ accordingly, the component description of the holomorphic three-form $\Omega$
of $X$ in (\ref{eq:OmegaX-BVCY-cmpnt}) can be brought into the form of 
\begin{align}
\Omega = 
\begin{pmatrix}
e''_j\wedge \hat{\alpha} & e''_{j+(22-r)/2}\wedge \hat{\alpha} & 
e''_{j}\wedge \hat{\beta} & e''_{j+(22-r)/2}\wedge \hat{\beta}
\end{pmatrix}
\begin{pmatrix}
\epsilon(\alpha_j) \\
\epsilon(\xi_S \alpha_j)\\
\tau \cdot \epsilon(\alpha_j)\\
\tau \cdot \epsilon(\xi_S \alpha_j)
\end{pmatrix},
\end{align}
where $j=1,\dots,(22-r)/2$. The $\Q$-basis of $V'_0 = T_0 \otimes H^1(E;\Q)$ employed above 
is now denoted by $\{ e_{i=1,\cdots, 2(22-r)} \}$. Using this basis, eigenvectors of $V'_0 \otimes \C$ 
diagonalizing the $K^S \times K^E$ algebra on $V'_0$ simultaneously are expressed as in 
\begin{align}
\begin{pmatrix}
\begin{pmatrix}
\Omega &
\Sigma_{a'} 
\end{pmatrix}&
\begin{pmatrix}
\Omegabar&
\Sigmabar_{a'}
\end{pmatrix}&
\Sigma'_a &
\Sigmabar'_a
\end{pmatrix}
=
\begin{pmatrix}
e_1 & \dots & e_{2(22-r)}
\end{pmatrix}
\begin{pmatrix}
M & M & M & M\\
\tau M & \taubar M & \tau M & \taubar M\\
\tau M & \taubar M & \taubar M & \tau M\\
\tau^2 M & \taubar^2 M & \abs{\tau}^2 M & \abs{\tau}^2 M
\end{pmatrix},
\end{align}
where 
\begin{align}
M:=
\begin{pmatrix}
\epsilon(\alpha_1)& \dots &\epsilon(\alpha_{(22-r)/2}) \\
\sigma_2(\alpha_1) &\dots &\sigma_2(\alpha_{(22-r)/2}) \\
\vdots & \ddots & \vdots \\
\sigma_{(22-r)/2}(\alpha_1)& \dots &\sigma_{(22-r)/2}(\alpha_{(22-r)/2}) \\
\end{pmatrix}, 
\end{align}
and $\sigma_{a=1,\cdots,(22-r)/2}$ are the embeddings $K^S_0 \hookrightarrow \R$.
A rationale behind this is as follows. 
First, let $\sigma^{\pm}_{a=1,\cdots,(22-r)/2}$ be the $(22-r)$ embeddings $K^S \hookrightarrow \C$, 
which satisfy $\sigma^{\pm}_a|_{K_0^S} = \sigma_a$ and $\sigma^{+}_a(\xi_S) = \tau$, $\sigma^-(\xi_S) = \bar{\tau}$; 
in this notation, $\epsilon = \sigma^+_1$ and $\bar{\epsilon} = \sigma^-_1$. Let $c^{\pm}$ be the 2 embeddings 
$K^E \hookrightarrow \C$, where $K^E= \Q( (\tau \times))$ is generated by a complex multiplication operation 
$(\tau \times)$, and $c^{\pm}$ maps $(\tau \times)$ to $\tau$ and $\bar{\tau}$, respectively. 
The $2(22-r)$ distinct embeddings of the algebra $K^S \times K^E$ into $\C$ are grouped into four, 
$(\sigma^+_a\cdot c^+)$, $(\sigma^-_a \cdot c^-)$, 
$(\sigma^+_a \cdot c^-)$ and $(\sigma^- \cdot c^+)$ with $a=1,\cdots, (22-r)/2$.  
The vectors $\Omega$ and $\Sigma_{a'=2,\cdots, (22-r)/2}$ in $V'_0 \otimes \C$ span eigenspaces of 
$K^S \times K^E$ corresponding to the embeddings $\sigma^+_a \cdot c^+$ 
($\Omega$ for $a=1$, and $a'=a$ otherwise). The three other groups of eigenvectors in $V'_0 \otimes \C$
correspond to the three other groups of embeddings in the order of appearance.

The last step is to exploit the fact that the complex structure $\tau$ of the elliptic curve $E$
has been assumed to have complex multiplication. Let us denote the minimal polynomial of $\tau$ as 
$\tau^2+p\tau+q$, with $p,q\in \Q$. Because 
\begin{align}
  \begin{pmatrix}
 p & 1 & 1 & 0 \\
 q & 0 & 0 & -1 \\
 0 & 1 & -1 & 0 \\
 q & p & 0 & 1 \\
\end{pmatrix} 
\begin{pmatrix}
1 & 1 & 1 & 1\\
\tau  & \taubar  & \tau  & \taubar \\
\tau  & \taubar  & \taubar  & \tau \\
\tau^2  & \taubar^2  & \abs{\tau}^2  & \abs{\tau}^2 
\end{pmatrix}
 = 
\begin{pmatrix}
\tau-\taubar & \taubar-\tau & 0 & 0\\
p\tau +2q & p\taubar +2q & 0 & 0\\
0& 0& \tau-\taubar & \taubar-\tau \\
0& 0&p\tau +2q & p\taubar +2q \\
\end{pmatrix}, 
\end{align}
a new basis of $V'_0$
\begin{align}
 \left( \underline{e}_1, \cdots, \underline{e}_{2(22-r)} \right) = 
 \left( e_1, \cdots, e_{2(22-r)} \right)
  \left(   \begin{pmatrix}
 p & 1 & 1 & 0 \\
 q & 0 & 0 & -1 \\
 0 & 1 & -1 & 0 \\
 q & p & 0 & 1 \\
\end{pmatrix} 
  ^{-1} \otimes {\bf 1}_{\frac{(22-r)}{2} \times \frac{(22-r)}{2}} \right)
\end{align}
allows us to split the vector space $V'_0$ over $\Q$ into 
$V_0 = {\rm Span}_\Q \left\{ \underline{e}_{i=1,\cdots, (22-r)} \right\}$ and 
$\underline{V}_0 = {\rm Span}_\Q \left\{ \underline{e}_{i=(22-r)+1,\cdots, 2(22-r)} \right\}$ 
so that 
\begin{align}
  V_0 \otimes \C = 
    {\rm Span}_\C \left\{ \Omega, \Sigma_{a'}, \overline{\Omega}, \overline{\Sigma}_{a'} \right\}, \qquad 
  \underline{V}_0 \otimes \C  = {\rm Span}_\C \left\{ \Sigma'_a, \overline{\Sigma}'_a \right\}.
\end{align}
$V_0$ is the level-3 component, and $\underline{V}_0$ the level-1.

The algebra $K^S \times K^E$ acts on $V_0$ and $\underline{V}_0$ separately, because we have 
already seen that the action of $K^S \times K^E$ can be diagonalized within $V_0 \otimes \C$ 
and $\underline{V}_0 \otimes \C$. Furthermore, the generator $(\tau \times)$ of $K^E$ over $\Q$
acts the same way as $\xi_S$ on $V_0$, and as $(-p-\xi_S)$ on $\underline{V}_0$. Therefore, 
the representation of the algebra $K^S \times K^E$ is not faithful on $V_0$ and $\underline{V}_0$; 
the algebra of Hodge endomorphisms on $V_0$ is $K^S$, and the same is true 
on $\underline{V}_0$. Under the assumption that the complex structure of the K3 surface $S$ is 
not in a Noether--Lefschetz locus of $D(T_0)$ (i.e., the rational Hodge structure on $T_0$ is 
simple, and $K^S$ is a field), the Hodge structure on $V_0$ and $\underline{V}_0$ are both 
simple, and moreover, $[K^S:\Q] = (22-r) = \dim_\Q V_0 = \dim_\Q \underline{V}_0$. 
We have now managed to keep the promise. 
 
Now, $(22-r)/2$ F-term conditions are implemented in $\underline{V}_0 \otimes \C$ and there are 
$(22-r)/2$ more F-term conditions in $V_0 \otimes \C$; the $\vev{W}=0$ condition is now purely on 
the $V_0 \otimes \C$ component. The analysis of F-term conditions in the level-1 
$\underline{V}_0 \otimes \C$ component is just the same as in section \ref{ssec:BV-just-F}. 
For a non-trivial flux to exist, so that the moduli of $S$ and $E$ are stabilized, it is 
necessary that 
\begin{align}
 (K^S)^r \cong \Q(\phi),
  \label{eq:cond-W0-1}
\end{align}
and the quadratic imaginary field $\Q(\phi)$ should be common to that for the twisted sectors. 
Moreover, 
\begin{align}
 \Q(\tau) \cong \Q(\phi)
  \label{eq:cond-W0-2}
\end{align}
because $K^S = K^S_0(\xi_S)$ and $\tau= \epsilon(\xi_S)$ should be in $\Q(\phi)$. 
$(22-r)$ NS--NS flux quanta can be chosen arbitrarily in this $\underline{V}_0$ component; 
the $(22-r)$ RR flux quanta are determined uniquely in order to reproduce a specified $\phi$ vev.

The same is true in the $V_0 \otimes \C$ component, if just the F-term conditions are imposed. 
The result on $\kappa$ in (\ref{eq:kappa-BV}) therefore does not change, even when the Hodge structure 
on $V'_0$ is not simple. For a vacuum to satisfy the condition $\vev{W}=0$, however, flux 
in the $V_0$ component should completely vanish. If there were a non-zero flux configuration in $V_0$, 
then the relation (\ref{eq:extra-linear-dep}) holds in $V_0$, which contradicts against the assumption
that the Hodge structure on $V_0$ (and also on $T_0$) is simple. Therefore, we come to the result 
\begin{align}
\kappa_0 =(22-r)+8g
 \label{eq:kappa0-BV}
\end{align}
for a complex structure $(z_S, \tau, \phi) \in D(T_0) \times {\cal H}_1 \times {\cal M}_{\rm dil.}$ 
satisfying the extra conditions (\ref{eq:cond-W0-1}, \ref{eq:cond-W0-2}), and 
\begin{align}
\kappa_0 = 0
\end{align}
otherwise. 

It is likely that all the moduli $z_S$, $\tau$ and $\phi$ are stabilized by a non-trivial flux 
in $\underline{V}_0$, when the conditions (\ref{eq:cond-W0-1}, \ref{eq:cond-W0-2}), even though 
there is no flux in $V_0$ so that the condition $\vev{W}=0$ is satisfied. To see this, 
note first that the quadratic order fluctuations of $z_S$ within $D(T_0)$ are contained 
also in the $(0,2)$ Hodge component of $T_0 \otimes \C$, and hence this quadratic fluctuations  
have overlap with the eigenspace $\C\overline{\Sigma}'_{a=1} \subset \underline{V}_0 \otimes \C$.
A non-trivial flux in the $(2,1)$ component of $\underline{V}_0 \otimes \C$ therefore 
gives rise to mass terms of the $z_S$ moduli. The chiral multiplets of $\tau$ moduli and $\phi$ 
moduli will have a Dirac mass term 
$\Delta W \propto - (m''_i \epsilon(y_i)) \; \phi \cdot (\delta \tau)$. This argument does not 
rule out a chance of accidental cancellation when all things considered, but such a cancellation 
is quite unlikely.\footnote{For a rigorous proof of non-zero masses without cancellation, we have 
to use a set of $(20-r)$ coordinates in a local patch of $D(T_0)$ around the vev $z_S \in D(T_0)$, to 
parametrize $\epsilon(y_i)$'s in (\ref{eq:SandE-period-def-cmpnt}). We do not do that in this article.} 
The flux configuration we have in mind here can be supersymmetric only when 
the moduli $(z_S, \tau, \phi)$ have very specific properties (encoded as very specific structure 
of the corresponding endomorphism fields); small deviation from such special loci would render 
the flux configuration non-supersymmetric, which is an indication that there is a scalar potential 
on the small deviation, the usual Noether--Lefschetz argument for the moduli stabilization by fluxes
in Type IIB string and F-theory. 

%%%%%%%%%%%%%%%%%%%%%%%%%%%%%%%%%%%%%%%%%%%%%%%%%%%%%%
\section{Non-Borcea--Voisin Cases}
\label{sec:non-BV}
%%%%%%%%%%%%%%%%%%%%%%%%%%%%%%%%%%%%%%%%%%%%%%%%%%%%%%

While we cannot have high hope of finding a series of infinite CM points in the complex 
structure moduli space of a family of Calabi--Yau threefolds other than in the form of 
Borcea--Voisin construction (Andr\'e--Oort conjecture), there are CM-type Calabi--Yau threefolds 
(e.g. some of Gepner models) isolated from other CM points. The analysis of the structure of 
endomorphism fields and the space of supersymmetric 
fluxes (\ref{eq:kappa-BV}, \ref{eq:kappa0-BV}) can be generalized easily for cases  
that are not Borcea--Voisin type. Let $X$ be such a Calabi--Yau threefold. 

First, when a complex structure $z \in {\cal M}_{\rm CM}^X \subset {\cal M}_{\rm cpx}^X$ is given, 
a rational Hodge structure is induced on the vector space $H^3(X;\Q)$. Suppose that 
\begin{align}
H^3(X,\Q)=V_0 \oplus V_1 \oplus \dots \oplus V_k
\end{align}
is a decomposition into simple rational Hodge substructures; $V_0$ is the level-3 
component, and all the others level-1 components. 
The algebra of Hodge endomorphisms of $V_a$ is then 
a CM field, denoted by $K^a$, and $[K^a: \Q] = \dim_\Q V_a$ for $a=0,\cdots, k$, as 
we assume a CM-type Calabi--Yau threefold $X$. 
The number field $K_{\rm tot}^X \subset \overline{\Q}$ is then the composite field 
of all the $(K^a)^{\rm nc}$'s in $\overline{\Q}$.

When we impose just the F-term conditions, analysis can be carried out (almost) 
separately for the individual components; among the $(1+h^{2,1}(X))$ F-term conditions 
of this theory, $\dim_\Q(V^a)/2$ of them are attributed to the $V_a$ component. 
For a non-trivial flux to be allowed in the $V_a$ component, we must have 
\begin{align}
  (K^a)^r \cong \Q(\phi)
  \label{eq:cond-F-1-general}
\end{align}
and the dilaton vev $\phi$ needs to generate a quadratic imaginary field;  
\begin{align}
 [\Q(\phi) : \Q] = 2.
  \label{eq:cond-F-2-general}
\end{align}
The reflex fields of all the CM fields $K^a$ should be one and the same quadratic 
imaginary field $\Q(\phi)$. This means, as in the previous section, that $\dim_\Q (V_a)=2$ 
and $K^a = \Q(\phi)$ for all of $a=1,\cdots, k$. The Weil intermediate Jacobian associated 
with the level-3 simple component $V_0$ must be isogenous to the product of 
$\dim_\Q(V_0)/2$ copies\footnote{All the level-1 simple components 
$V_a$ with $a \geq 1$ are of 2-dimensions over $\Q$, but the level-3 simple component 
$V_0$ is not necessarily of 2-dimensions (see also footnote \ref{fn:cycl-8}).}
of an elliptic curve with complex multiplication in $\Q(\phi)$.
It thus follows that the Weil intermediate Jacobian associated with 
$H^3(X;\Q)$ needs to be isogenous to $(E_\phi)^{b_3/2}$. 
The space of fluxes consistent with the F-term conditions at this complex structure 
$\vev{z,\phi}$ has a dimension 
\begin{align}
  \kappa = \sum_{a=0}^k \dim_\Q V_a = b_3 (X).
 \label{eq:kappa-general-A}
\end{align}

Not all the flux configurations in a $\kappa$-dimensional space over $\Q$
end up with a vacuum with $\vev{W} = 0$; the necessary and sufficient condition for 
$\vev{W}=0$ is that the NS--NS and RR flux vanish in the level-3 simple component $V_0$. 
Therefore, the flux configurations satisfying both the F-term conditions and $\vev{W}=0$ 
form a subspace with a dimension 
\begin{align}
 \kappa_0 = \sum_{a=1}^k \dim_\Q V_a. %  = b_3(X) - \dim_\Q V_0.
  \label{eq:kappa0-general-A}
\end{align}
Note that the condition $(K^{a})^r \cong \Q(\phi)$ does not have to be satisfied 
for the level-3 $(a=0)$ component for such a non-trivial supersymmetric flux configuration 
to exist. We should also remind ourselves, however, that there is no guarantee whether all the 
moduli in ${\cal M}_{\rm cpx}^X \times {\cal M}_{\rm dil.}$ are given supersymmetric masses 
in this situation. 

Four examples of non-Borcea--Voisin CM-type Calabi--Yau threefolds $M_m$ ($m=5,6,8,10$) 
are studied in section 5 of \cite{DeWolfe}. They are the Gepner points in the complex 
structure moduli spaces of four families of Calabi--Yau threefolds, all with $h^{2,1}=1$.
The field $K_{\rm tot}$ in the four examples are all cyclotomic fields, $\Q(\zeta_m)$, 
with $m=5,6,8,10$, respectively. 
The characterization (\ref{eq:cond-F-1-general}, \ref{eq:cond-F-2-general}) of the 
number fields of Hodge decomposition and the formula of the space of 
supersymmetric fluxes (\ref{eq:kappa-general-A}, \ref{eq:kappa0-general-A}) reproduce 
all the results for the four examples obtained in \cite{DeWolfe} in a systematic way.

Let us first look at the two examples $M_m$ with $m=5, 10$. 
The cohomology group $H^3(X;\Q)$ as a whole forms a simple rational Hodge structure 
in the two cases; in the CM type $(K, \Phi)$ of the Weil intermediate Jacobian 
of $H^3(M_m;\Q)$, the endomorphism field $K$ is generated by the minimum phase 
$+2\pi$ monodromy $\theta_m$ around the corresponding Gepner point of the complex structure 
moduli space, $K \cong \Q(\theta_m) \cong \Q(\zeta_m)$, and $\Phi$ consists\footnote{
For a cyclotomic field $K = \Q(\theta_m) \cong \Q(\zeta_m)$, the $\varphi(m)$ embeddings 
of $K$ to $\C$ are denoted by $\phi_j$, $j \in (\Z/(m\Z))^\times$; 
$\phi_j: \theta_m \mapsto (\zeta_m)^j$. } of $\{\phi_1, \phi_{3}\}$ in the $m=5$ case and 
of $\{ \phi_1, \phi_{7} \}$ in the $m=10$ case. The reflex field remains 
$\Q(\zeta_m) \subset \overline{\Q}$ in those 
two cases (e.g., Ex. 3.2. (c), \cite{kerr2010shimura}), which means that 
the condition (\ref{eq:cond-F-1-general}, \ref{eq:cond-F-2-general}) cannot be satisfied 
regardless of the value of $\phi$. Therefore $\kappa = \kappa_0 = 0$.

The example $M_{m=8}$ is a little different from the examples with $m=5,10$. 
Its Weil intermediate Jacobian introduces a weight-1 rational Hodge structure on 
the $4$-dimensional vector space $H^3(M_m;\Q)$; in the CM type $(K, \Phi)$, 
the endomorphism field is generated by the minimum $+2\pi$ monodromy $\theta_m$ 
around the Gepner point, $K = \Q(\theta_m) \cong \Q(\zeta_m)$, and $\Phi = \{\phi_1, \phi_5 \}$.
This CM type is not primitive, but induced from a CM type on a subfield 
$\Q(\sqrt{-1}) \subset K$. The reflex field of these CM types is 
$\Q(\sqrt{-1}) \subset \overline{\Q}$, which is a quadratic imaginary field. 
Therefore, $\kappa = 4$ if the dilaton vev satisfies $\phi \in \Q(\sqrt{-1})$; 
$\kappa = 0$ otherwise.
There is no non-trivial flux configuration satisfying $\vev{W}=0$ at this Gepner point (i.e., 
$\kappa_0 = 0$), however, regardless of the value of $\phi$; 
this is because the weight-3 rational Hodge structure on $H^3(M_m;\Q) \cong V_0$ 
is simple.\footnote{\label{fn:cycl-8}The weight-1 rational Hodge 
structure of the Weil and Griffiths intermediate Jacobians are not simple, but 
the weight-3 rational Hodge structure on $H^3(M_m;\Q)$ is simple in this example.}

In the example $M_{m=6}$, on the other hand, the weight-3 rational Hodge structure 
on $H^3(M_m;\Q)$ is not simple. $H^3(M_m;\Q)$ can be split into two 2-dimensional 
subspaces $V_0 \oplus V_1$ on which the rational Hodge 
substructures are level-3 and level-1, respectively. The endomorphism field of those two components are both 
$K = \Q(\theta_m) \cong \Q(\zeta_m)$ generated by the minimum phase $+2\pi$ monodromy 
$\theta_m$ around the Gepner point; now $\dim_\Q (V_a) = 2 = [K:\Q]$, and hence $M_{m=6}$ is a CM-type 
Calabi--Yau. Now, $(K_{\rm tot}^{M_m})^r = K_{\rm tot}^{M_m} = \Q(\zeta_6) \cong \Q(\sqrt{-3})$. 
So, $\kappa = 4$ and $\kappa_0 = 2$ if $\phi \in \Q(\sqrt{-3})$. 
If $\phi \nin \Q(\sqrt{-3})$, however, $\kappa = \kappa_0 = 0$.

Our results for the four $M_m$'s, derived in this section, perfectly agree with those in \cite{DeWolfe}.

%%%%%%%%%%%%%%%%%%%%%%%%%%%%%%%%%%%%%%%%%%%%%%%%%%%%%%
\section{Discussions}
\label{sec:discussions}
%%%%%%%%%%%%%%%%%%%%%%%%%%%%%%%%%%%%%%%%%%%%%%%%%%%%
There are a few things to remark after carrying out the analysis in the preceding 
sections. 

\begin{anythng}
\label{statmnt:fluxvac-enrichmnt}
It turns out that the variety of supersymmetric flux configurations 
at a CM point in ${\cal M}_{\rm CM} \subset {\cal M}_{\rm alg} \subset {\cal M}_{\rm cpx}^X 
\times {\cal M}_{\rm dil.} $ is quite different from what is expected under the general 
formula (\ref{eq:dW-formula-F}, \ref{eq:dW-formula-W}) for points in ${\cal M}_{\rm alg}$. 
The number field $K_{\rm tot}$ obtained from the CM fields of simple Hodge substructures 
may have an extension degree $d_{K{\rm tot}} = [K_{\rm tot}:\Q]$ larger than that
of the individual CM fields, first of all. Secondly, the dimension $\kappa$ [resp. $\kappa_0$] 
of the $\Q$-vector space of supersymmetric fluxes can be much larger than 
the expectation (\ref{eq:dW-formula-F}, \ref{eq:dW-formula-W}), when the CM fields satisfy 
extra conditions (\ref{eq:cond-F-1-BV}, \ref{eq:cond-F-2-BV}, \ref{eq:cond-F-3-BV}) 
[resp.  (\ref{eq:cond-F-1-BV}, \ref{eq:cond-F-2-BV}, \ref{eq:cond-F-3-BV}, \ref{eq:cond-W0-1},
\ref{eq:cond-W0-2})] in the Borcea--Voisin case, 
and (\ref{eq:cond-F-1-general}, \ref{eq:cond-F-2-general})
[resp. (\ref{eq:cond-F-1-general}, \ref{eq:cond-F-2-general}) except
(\ref{eq:cond-F-1-general}) for $a=0$] in non-Borcea--Voisin cases. 
When those extra conditions are not satisfied, on the other hand, the space of 
supersymmetric fluxes just becomes trivial, $\kappa = 0$ / $\kappa_0 = 0$. 

This discrepancy between the cases in ${\cal M}_{\rm alg}$ and ${\cal M}_{\rm CM}$ is because 
the general expectation in (\ref{eq:dW-formula-F}) and (\ref{eq:dW-formula-W}) was 
derived based on the assumption that all the F-term conditions (and 
also the $\vev{W}=0$ condition) give rise to the conditions on the flux quanta that are 
mutually linearly independent over $\Q$ \cite{DeWolfe}.
We have seen that those conditions of supersymmetric fluxes are far from mutually 
independent for any point in ${\cal M}_{\rm CM}$, as have also been observed in the example $M_8$
of \cite{DeWolfe}. The key observation is the structure of eigenvectors (\ref{eq:key-structure}) that 
supports the Hodge decomposition at a CM point. In a given component of a simple Hodge substructure $V_a$, there is 
essentially just one F-term condition in $K^a$, or $[K^a:\Q] =: d_{A^a}$ conditions independent over $\Q$. Therefore, a 
modified version $\kappa = 2b_3 - \sum_{a=0}^k d_{K^a} \times (\dim_\Q V_a)/2$ of (\ref{eq:dW-formula-F}) for a generic point in ${\cal M}_{\rm alg}$ should be further replaced by   
\begin{align}
\kappa=2b_3 - \sum_{a=0}^k d_{K^a} \times 1 = 2b_3 - \sum_{a=0}^k (\dim_\Q V_a) = b_3
\end{align}
for a CM point, which was the essence behind the results (\ref{eq:kappa-BV}, \ref{eq:kappa-general-A}). 
The formula $\kappa_0 = \kappa - d_{K{\rm tot}}$ in (\ref{eq:dW-formula-W}) is modified 
to 
\begin{align}
\kappa_0= \kappa - d_{K^{a=0}}  % b_3 - \dim_\Q V_0=\sum_{i>0} \dim_\Q V_i.
\end{align}
to be the version appropriate for a CM-type Calabi--Yau threefold.\footnote{
It is also worth reminding ourselves that the rational Hodge structure on $H^3(X;\Q)$ 
cannot be simple for a non-trivial flux vacuum to exist on a CM-type Calabi--Yau 
threefold $X$; see (\ref{eq:extra-linear-dep}). This observation itself is not entirely new, however.}
\end{anythng}
 
\begin{anythng}
For a non-trivial supersymmetric flux configuration to exist on a CM-type Calabi--Yau 
threefold $X$, it turns out that the endomorphism fields $K^a$ of all the simple 
components of the rational Hodge structure of $H^3(X;\Q)$ have the reflex field isomorphic 
to a common quadratic imaginary field $\Q(\phi)$, which is generated by the dilaton 
vev $\phi$. Its immediate consequence is that the Abelian variety associated with 
the level-1 simple components $V_a$ (for $a=1,\cdots, k$) are elliptic curves isogenous 
to $E_\phi$, and defined over a number field that is an Abelian extension over $\Q(\phi)$. 
If we introduce fluxes also in the $V_0$ component (by giving up the $\vev{W}=0$ 
condition), then the Weil intermediate Jacobian can also be defined over an Abelian extension 
of $\Q(\phi)$.
\end{anythng}

It is also striking that the conditions for existence of non-trivial supersymmetric 
flux configurations (\ref{eq:cond-F-1-general}) are stated in terms of reflex fields, 
because a zero-dimensional Shimura variety, a collection of CM points, is specified 
for a Mumford--Tate group that is an image of the multiplicative group of the reflex 
field (\ref{eq:MT-reflex-rltn}). We can therefore think of classifying CM points, 
preliminary, by their quadratic imaginary reflex field $\Q(\phi)$, and then by 
embeddings of $\Q(\phi)^\times \rightarrow \mathbb{G}{\rm Sp}(b_3)$. Hard math problems, 
such as a) and b) in the case of Borcea--Voisin Calabi--Yau threefolds and the 
alignment of ${\cal M}_{\rm cpx}^X$ with those CM points in the non-Borcea--Voisin cases, 
should be imposed on top of this preliminary classifications, however.

\begin{anythng}
Points in ${\cal M}_{\rm alg}^X$ that are not within its subset ${\cal M}_{\rm CM}^X$ do not need 
to be taken out of picture in the original idea in \cite{DeWolfe}. As reviewed briefly 
in section \ref{sec:math-bkgd} in this article, CM points ${\cal M}_{\rm CM}^X$ forms a much smaller subset 
of ${\cal M}_{\rm alg}^X$. While there are so numerous points in ${\cal M}_{\rm alg}^X$, however, 
the number of supersymmetric flux vacua should still be estimated by (\ref{eq:dW-formula-F}, 
\ref{eq:dW-formula-W}) for those that are not in ${\cal M}_{\rm CM}^X$, and there can often 
be no supersymmetric flux when $[K_{\rm tot.} : \Q]$ is moderately large. For CM points where 
the reflex field is $\Q(\phi)$ that is quadratic imaginary, however, much greater number 
of supersymmetric flux configurations are allowed, essentially due to the special 
structure (\ref{eq:key-structure}) in the Hodge decomposition at CM 
points (\ref{statmnt:fluxvac-enrichmnt}). Due to this enrichment of flux vacua on CM points, 
it is not obvious which side wins in the game of flux vacua counting, CM points, or non-CM 
algebraic points.  
\end{anythng}

In the context of flux vacua counting, one will also be interested in the fact that the 
condition (\ref{eq:cond-F-1-general}) for the level-3 simple component $V_0$ does 
not have to be imposed, when we seek for flux vacua with $\vev{W}=0$. This means 
that a larger subset of CM points than those satisfying (\ref{eq:cond-F-1-general}) 
admit non-trivial supersymmetric flux configurations in all the level-1 simple components, 
but this opportunity comes with a chance that some of moduli in ${\cal M}_{\rm cpx}^X$ 
remain unstabilized; if we set the flux in the level-3 component $V_0'$ to be trivial 
in section \ref{ssec:BV-just-F}, at least the moduli $\tau \in {\cal M}_{\rm cpx}^{\rm ell.}$
would have remained massless, for example. We did not encounter a single example in this 
article (within Borcea--Voisin or $M_{m}$'s) where i) all the moduli in 
${\cal M}_{\rm cpx}^X \times {\cal M}_{\rm dil.}$ are stabilized without a flux in the 
simple level-3 component $V_0$, and ii) the condition (\ref{eq:cond-F-1-general}) 
is not satisfied for the $a=0$ component at the same time. It is an open curious 
math question whether and how many such CM points exist. 

\begin{anythng}
A greater problem still is if there is any reason to pay attention only to the vacua within 
${\cal M}_{\rm alg}^X \subset {\cal M}_{\rm cpx}^X$ and forget others in the first place. 
As texts in section \ref{sec:math-bkgd} already clarify the way we think, we just do not have 
any justification based on the current understanding of string theory. Justification may come 
from developments on string theory in the future, such as consistency analysis of fluxes in 
world-sheet language. Or possibly string theory may just have to be formulated only for 
$(c,c)$ rings that sit on CM points. If that is the case, then one would not have to bother 
about the statistics game above in the first place. 
\end{anythng}

\begin{anythng}
Obviously the presence of dilaton chiral multiplet plays essential roles in stating the 
result of flux analysis on CM-type Calabi--Yau threefolds. It is therefore an interesting 
question how the analysis should be modified, when we think of F-theory compactifications 
on CM-type Calabi--Yau fourfolds. Also, possible enrichment of flux vacua on loci with discrete 
symmetries in such an arithmetic regime was also a question of interest in 
\cite{DenefDist,DeWolfe}. That will be all the more interesting question, when studied in 
F-theory compactifications. We therefore leave the study involving F-theory to our future works. 
\end{anythng}

%%%%%%%%%%%%%%%%%%%%%%%%%%%%%%%%%%%%%%%%%%%%%%%%%%%%
\subsection*{Acknowledgements}  % in BrE
% \subsection*{Acknowledgments}   % in AmE
%%%%%%%%%%%%%%%%%%%%%%%%%%%%%%%%%%%%%%%%%%%%%%%%%%%%%

We thank T. Abe and S. Kondo for tutoring us in basics of number theory, M. Weissenbacher 
for discussions, and N. Yui for providing us with a copy of \cite{yui2013modularity}, 
from which our exploration started. TW thanks Harvard theory group for hospitality, 
where the earliest stage discussions of this work took place. This work is supported 
in part by Advanced Leading Graduate Course for Photon Science grant (KK), 
Grant in Aid no. 26287039 (KK), Brain circulation program (TW) and 
the WPI Initiative (KK and TW), MEXT, Japan. 

\appendix 

%%%%%%%%%%%%%%%%%%%%%%%%%%%%%%%%%%%%%%%%%%%%%%%%%%%%%
\section{Field theory}
\label{sec:field}
%%%%%%%%%%%%%%%%%%%%%%%%%%%%%%%%%%%%%%%%%%%%%%%%%%%%%

Anything in the appendix \ref{sec:field} should be found in standard textbooks on 
field theory; most of materials in the appendix \ref{sec:Hodge} are also well-known 
facts written down explicitly (or used implicitly) in the literatures.
Our primary sources are \cite{roman2005field,Fujisaki} for \ref{ssec:field_basics}
and \cite{huybrechts2016lectures,shimura1961complex,shimura2016abelian,ShimuraIntro,borceacm}
for \ref{ssec:field_cm}--\ref{ssec:reltn-Hdg-cmp-embd}
(see also footnotes \ref{fn:math-sources} and \ref{fn:math-sources2}).
Those materials are placed here in the preprint version for convenience of the readers. 
They may thus be dropped from a journal version, following suggestions from referees 
and the editor.  

%%%%%%%%%%%%%%%%%%%%%%%%%%%%%%%%%%%%%%%%%%%
\subsection{Basics}
\label{ssec:field_basics}
%%%%%%%%%%%%%%%%%%%%%%%%%%%%%%%%%%%%%%%%%%%

In this article, it is always assumed that fields have characteristic zero, and 
a commutative multiplication law.

%%%%%%%%%%%%%%%%%%%%%%%%%%%%%%%%%%%%%%%%%%%%%%%
\subsubsection{Algebraic Extension and Algebraic Number}
%%%%%%%%%%%%%%%%%%%%%%%%%%%%%%%%%%%%%%%%%%%%%%%

\begin{defn}
A nonempty subset $F$ of a field $E$ is called a \dfn{subfield} of $E$ if it is a field with the same operations as in $E$. \\
If $F$ is a subfield of a field $E$, we call $E$ an \dfn{extension field} of $F$. This \dfn{extension} is denoted by $E/F$.
\end{defn}

\begin{defn}
Let $E/F$ be an extension and $S$ be a subset of $E$. The smallest subfield of $E$ containing both $F$ and $S$ is denoted by $F(S)$. If $S=\b{\alpha_1,\dots,\alpha_n}$ is a finite set, then the extension $F(S)/F$ is said to be \dfn{finitely generated} and denoted by $F(\alpha_1,\dots,\alpha_n)$. An extension of the form $F(\alpha)/F$ is said to be \dfn{simple} and $\alpha$ is called a \dfn{primitive element}.
\end{defn}

\begin{defn}
Let $F$ be a field and $E$ be an extension field of $F$. If an element $x \in E$ is a root of some polynomial with all the coefficients in $F$, then $x$ is said to be \dfn{algebraic} over $F$. Otherwise $x$ is said to be \dfn{transcendental} over $F$.
An extension $E/F$ is called an \dfn{algebraic extension} if every element in $E$ is algebraic over $F$.
\end{defn}

\begin{defn}
An extension field $E$ of a field $F$ can be viewed as a vector space over $F$. The dimension of the vector space is called the \dfn{degree} of the extension and denoted by $[E:F]$. If $[E:F]$ is finite, then $E/F$ is called a \dfn{finite extension}.
\end{defn}

\begin{thm}
\label{thm:fin-ext=fin-alg}
Let a field $K$ be an extension field of a field $F$. Then the following conditions are equivalent:
\begin{enumerate}
\item
$K$ is a finite extension field of $F$, i.e., $[K:F]<\infty$.
\item
$K$ is a finitely generated algebraic extension field of $F$.
\end{enumerate}
\end{thm}

\begin{thm}\label{thm:extdeg}
Let $E/F$ and $K/E$ be extensions. Then
\begin{align}
[K:F]=[K:E][E:F].
\end{align}
If $\{\alpha_i \mid i=1,\dots,[E:F]\}$ is a basis of the vector space $E$ over $F$, and 
$\{\beta_j \mid j=1,\dots,[K:E]\}$ that of $K$ regarded as a vector space over $E$, then the set of products 
$
\{\alpha_i \beta_j \mid i=1,\dots,[E:F],\; j=1,\dots,[K:E]\}
$
is a basis of the vector space $K$ over $F$.
\end{thm}

\begin{center}
  ........................................................
\end{center}

\begin{defn}
For a field $F$, $F[x]$ denotes the ring of polynomials in a single variable $x$ with all the coefficients 
in $F$. For a finite algebraic extension $E/F$, and for an element $\alpha \in E$, non-zero polynomials 
$p_\alpha(x) \in F[x]$ satisfying $p_\alpha(\alpha) = 0 \in E$ with the smallest degree possible are called 
\dfn{minimal polynomials of $\alpha$ over} $F$. Such a polynomial always exist (because $\alpha$ is 
algebraic over $F$), and is unique up to overall multiplication of elements in $F^\times$. 
Minimal polynomials are always irreducible in $F[x]$.
\end{defn}

\begin{thm}
Let $K/F$ be an extension and let $\alpha\in K$ be algebraic over $F$. Then the subfield 
$F(\alpha)$ of $K$ has a structure
\begin{align}
F(\alpha) \simeq F[x]/(p_\alpha(x)),
\end{align}
where $p_\alpha$ is a minimal polynomial of $\alpha$ over $F$.
\end{thm}

In fact, a finite algebraic extension $K/F$---not just a subfield $K(\alpha) \subset K$---always 
has a structure like that, when ${\rm char}(F) = 0$; this useful property is stated as follows:

\begin{lemma}
\label{lmma:smpl-ext}
When ${\rm char}(F) =0$, any finite extension $K/F$ is a simple extension; that is, there exists an 
element $\theta \in K$ so that $K = F(\theta)$. 
Using a minimal polynomial of $\theta$ over $F$, therefore, the field $K$ has a structure 
$K \cong F[x]/(p_\theta(x))$. It is always possible to take $\{ 1, \theta, \theta^2, \cdots, \theta^{[K:F]-1}\}$
as a basis, when $K$ is regarded as a $[K:F]$-dimensional vector space over $F$. 
\end{lemma}

\begin{exmpl}
This theorem states that even a field that is generated by multiple elements can be thought of as a simple extension. For example,
$\Q(i\sqrt{2},i\sqrt{3})=\Q(i\sqrt{2}+i\sqrt{3})$.
\end{exmpl}

\begin{center}
  ........................................................
\end{center}

All the definitions and theorems on algebraic extension so far are for fields that 
are defined abstractly by the laws of addition and multiplication among their elements.
We may sometimes have a little more specific interest, however, in a field $K$ that is defined as 
a subfield of $\C$. For such a field $K$, ${\rm char}(K)=0$ by definition.

\begin{defn}
A complex number $\alpha \in \C$ is called an \dfn{algebraic number}, if there is 
a non-zero polynomial $p_\alpha(x) \in \Q[x]$ satisfying $p_\alpha(x=\alpha) = 0 \in \C$.
It is known that all the algebraic numbers form a subfield of $\C$; this subfield 
is denoted by $\overline{\Q}$. Any finite extension field $K$ of $\Q$ that is defined 
as a subfield of $\C$ is called a \dfn{number field}. 
\end{defn}

Any number field is always a subfield of $\overline{\Q}$. 
While $\Qbar/\Q$ is an algebraic extension, it is not a finite extension.
Thus, $\Qbar$ itself is not a number field.
%While all the elements of $\overline{\Q}$ are algebraic over $\Q$, 
%$\overline{\Q}/\Q$ is not a finite extension. 

%%%%%%%%%%%%%%%%%%%%%%%%%%%%%%%%%%%%%%%%%%%%%%%%%%%%%%%
\subsubsection{Embeddings into $\C$}
%%%%%%%%%%%%%%%%%%%%%%%%%%%%%%%%%%%%%%%%%%%%%%%%%%%%%%

Here is a summary of results on embeddings of a finite extension field $K$ over $\Q$
into a subfield of $\C$. We begin, however, with the following preparation.

\begin{thm}
Let $K$ be an algebraic extension over $\Q$, and $\alpha \in K$. For a minimal polynomial 
$p_\alpha(x)$ of $\alpha$ over $\Q$, there are ${\rm deg}(p_\alpha)$ solutions to $p_\alpha(x) = 0$ in $\C$. 
It is known that all the roots of $p_\alpha(x) = 0$ come with multiplicity 1. 
\end{thm}

This property is valid for any algebraic extension over $F$, in place of $\Q$, as long as ${\rm char}(F)=0$,
and is called \dfn{separability}.

\begin{thm}\label{thm:dist-embd}
Let $K$ be a finite extension over $\Q$. Then there are $[K:\Q]$ distinct embeddings 
(isomorphism onto the image) $\rho:K \hookrightarrow \C$ over $\Q$. Since all the elements 
in $K$ are algebraic, the image of such an embedding is always contained within $\overline{\Q}$; 
$\rho(K) \subset \overline{\Q} \subset \C$.   
\end{thm}

This is because $K$ can always be regarded as a simple extension over $\Q$ by a primitive element 
$\theta \in K$ (Lemma \ref{lmma:smpl-ext}); let $p_\theta(x)$ be its minimal polynomial over $\Q$, 
and $\{ \xi_{i = 1,\cdots, [K:\Q]}\} \subset \C$ be the roots of $p_\theta(x)=0$ in $\C$.
Then $\rho_i: K \hookrightarrow \C$ is given by 
$\rho_i: K \ni \theta \mapsto \rho_i(\theta) = \xi_i \in \C$ for $i=1,\cdots, [K:\Q]$.
Note that all the $[K:\Q]$ roots $\{\xi_i\}$ are distinct from one another (separable), and hence 
the corresponding embeddings are distinct from one another. 

\begin{center}
  ............................................
\end{center}

\begin{defn}
Now let $K/F$ be a finite extension with degree $m =[K:F]$. For any element $x \in K$, then, 
$A(x): y \mapsto x \cdot y$ for $y \in K$ is an $F$-linear transformation on the 
vector space $K$ over $F$. ${\rm Tr}_{K/F}(x)$ denotes the trace of the $F$-valued $m \times m$ matrix 
representation of $A(x)$, and is called the {\bf trace} of $x \in K$.
\end{defn}

\begin{anythng}
Let $\{ \omega_{i=1,\cdots, m} \}$ be a basis of $K$ as a vector space over $F$. Then 
\begin{align}
  x \cdot \omega_i = \omega_j [A(x)]_{ji},
 \label{eq:multiply-reltn}
\end{align}
where $[A(x)]_{ji}$ is the $F$-valued $m \times m$ matrix representation of $A(x)$.
Now, let us take $F = \Q$. The relation (\ref{eq:multiply-reltn}) among elements in $K$ still holds 
as one among their images under the embeddings of $K$ into $\overline{\Q} \subset \C$. 
\begin{align}
 \rho_a(x) \rho_a(\omega_i) = \rho_a(\omega_j) [A(x)]_{ji}.
  \label{eq:multiply-reltn-embdd}
\end{align}
Since there are $m$ distinct embeddings $\rho_{a=1,\cdots, m}: K \rightarrow \overline{\Q} \subset \C$, 
$\rho_{a}(\omega_i)$, $\rho_a(\omega_j)$ and $\rho_a(x)$ can be regarded as $\C$-valued $m \times m$ 
matrices (the matrix $\rho_a(x)$ is diagonal), and the following relation is obtained:
\begin{align}
 {\rm Tr}_{K/\Q}(x) = \tr_{m \times m}\left[ A(x) \right] = \sum_{a=1}^m \rho_a(x);
\end{align}
each contribution on the right-hand side is an algebraic number in $\overline{\Q} \subset \C$, 
but their sum should be in $\Q$, because the left-hand side is, by definition. 
\end{anythng}

%%%%%%%%%%%%%%%%%%%%%%%%%%%%%%%%%%%%%%%%%%%%%%%%%%%%%%%%%%%%%%%%%
\subsubsection{Normal Closure}
%%%%%%%%%%%%%%%%%%%%%%%%%%%%%%%%%%%%%%%%%%%%%%%%%%%%%%%%%%%%%%%%%

\begin{defn}
Let $K$ be a number field, i.e., a subfield of $\overline{\Q} \subset \C$ that is a
finite extension over $\Q$. Let $\theta$ be a primitive element (i.e., $K =\Q(\theta)$), 
$p_\theta(x)$ be its minimal polynomial over $\Q$, and $\{ \xi_1 = \theta, \xi_2, \cdots, \xi_{[K:\Q]}\}$ 
be the roots of $p_\theta(x) = 0$ in $\C$. The field $\Q(\xi_1,\cdots, \xi_{[K:\Q]}) \subset \overline{\Q}$ 
is called the \dfn{smallest splitting field of $p_\theta(x) \in \Q[x]$ in $\overline{\Q}$}. 
\end{defn}

\begin{anythng}
Thinking of a number field $K$ as an abstract finite extension field over $\Q$, we see 
that there must be $[K:\Q]$ embeddings $\rho_i: K \hookrightarrow \overline{\Q} \subset \C$, 
$i=1,\cdots, [K:\Q]$ (Thm \ref{thm:dist-embd}). The embedding $\rho_{i=1} : K \hookrightarrow \overline{\Q}$ is a trivial 
identification, and $\rho_1(K) = K \subset \overline{\Q}$. For other $\rho_i$'s, however, it is not 
guaranteed that $\rho_i(K) = K$.  
\end{anythng}

\begin{anythng}
The field $\Q(\xi_1, \cdots, \xi_{[K:\Q]})$ can be regarded as the minimal subfield of $\overline{\Q}$ 
that contains all the images $\cup_{i=1,\cdots, [K:\Q]}\rho_i(K)$ of the $[K:\Q]$ embeddings from $K$ to 
$\overline{\Q}$. Because of this characterization, the smallest splitting field 
$\Q(\xi_1,\cdots, \xi_{[K:\Q]}) \subset \overline{\Q}$ of $p_\theta(x)$ in $\overline{\Q}$ does not 
depend on the choice of a primitive element $\theta$. 
\end{anythng}

\begin{thm}\label{thm:nc}
For a subfield $K^{\rm nc} := \Q(\xi_1,\cdots, \xi_{[K:\Q]})$ of $\overline{\Q}$ for a number field $K$, 
any one of the embeddings $\rho: K^{\rm nc} \hookrightarrow \overline{\Q}$ over $\Q$ maps $K^{\rm nc}$ 
to $K^{\rm nc} \subset \overline{\Q}$, not outside of $K^{\rm nc}$ (though not necessarily as a trivial 
map on $K^{\rm nc}$)---(*). This is because such an embedding $\rho$ has to send $\xi_i$'s to $\xi_i$'s,
possibly with a permutation among them, and cannot do anything more than that. 
\end{thm}

\begin{anythng}\label{statmnt:nc2}
A subfield $E$ of $\C$ is said to be a \dfn{normal extension of} $\Q$, if it has the property (*) 
referred to above. The minimum subfield in $\C$ of a number field $K$ that is a normal extension over 
$\Q$ is called the \dfn{normal closure of $K/\Q$ in} $\C$, and is denoted by $K^{\rm nc}$, as we have done already above. 
For a number field $K$, therefore, the smallest splitting field 
$\Q(\xi_1, \cdots, \xi_{[K:\Q]}) \subset \overline{\Q}$ of a primitive element $\theta$ such that 
$K=\Q(\theta)$ is the normal closure of $K$. 

For a finite extension field $E$ over $\Q$ that is defined as an abstract field, one can pick 
any one of embeddings $\rho: E \hookrightarrow \overline{\Q} \subset \C$. The normal closure 
of $\rho(E)$ in $\overline{\Q}$ does not depend on which one of $[E:\Q]$ embeddings is used.
So, we use a notation $E^{\rm nc}$ for $(\rho(E))^{\rm nc}$ in this article.
\end{anythng}

The definition of a normal extension $E\subset \Qbar$ over $\Q$ is generalized to extensions
$E\subset \Qbar$ over an arbitrary number field $F$ by replacing $\Q$ in \ref{thm:nc} and \ref{statmnt:nc2} with $F$.

\begin{defn}
An algebraic extension $E/F$ is said to be \dfn{Galois}, if it is a separable and normal extension. 
Note that the separability is always guaranteed, when $F$ has ${\rm char}(F)=0$. 
\end{defn}
 
\begin{exmpl}
The normal closure $K^{\rm nc}$ of a number field $K$ is always a Galois extension over $\Q$, by definition. 
Not all the number fields $K$ are Galois over $\Q$, however. Cyclotomic fields $K = \Q(\zeta_N)$ are 
examples of Galois extensions over $\Q$; the number field $K = \Q[x]/(x^3-2)$, on the other hand,
is not Galois; another example of non-Galois extension is found in Example \ref{exmpl:quartic}.
\end{exmpl}

%%%%%%%%%%%%%%%%%%%%%%%%%%%%%%%%%%%%%%%%%%%%%%%%%%%%%%%%%
\subsection{CM Field}
\label{ssec:field_cm}
%%%%%%%%%%%%%%%%%%%%%%%%%%%%%%%%%%%%%%%%%%%%%%%%%%%%%%%%%%

First, we prepare a few jargons.

\begin{defn}
A finite extension field $K$ over $\Q$ is said to be \dfn{totally real} if $\rho_i(K) \subset \R$ 
for all the $[K:\Q]$ embeddings $\rho_{i=1,\cdots, [K:\Q]}: K \hookrightarrow \C$.
On the other hand, a finite extension field $K$ over $\Q$ is said to be \dfn{totally imaginary} 
if $\rho_i(K)$ is not contained within $\R$ for any one of the $[K:\Q]$ embeddings 
$\rho_{i=1,\cdots, [K:\Q]}: K \hookrightarrow \C$. 
\end{defn}

\begin{exmpl}
Let $n \in \Z$ and suppose that $\abs{n}$ is not the square of an integer.
$K=\Q[x]/(x^2-n)$ is totally real [resp. totally imaginary]
if $n > 0$ [resp. $n < 0$]. This field $K$ has two embeddings into $\C$; 
$\rho_{\pm}: x \mapsto \pm \sqrt{n} \in \C$. On the other hand, $K=\Q[x]/(x^3-2)$ is neither totally 
real nor totally imaginary; the three embeddings of $K$ to $\C$ send $x \in K$ to one of the 
three roots of $x^3-2=0$ in $\C$. 
\end{exmpl}

Now, here is the definition of a CM field.

\begin{defn}
A finite extension field $K$ over $\Q$ is said to be a \dfn{CM field}, if (i) it contains 
a subfield $K_0$ that is totally real, (ii) $K$ is a degree-2 extension of $K_0$, and (iii)
$K$ itself is totally imaginary. Therefore, $[K:\Q] = [K:K_0][K_0:\Q]=2[K_0:\Q]$ is always 
an even integer. 
\end{defn}

\begin{props}
\label{props:CM-embd-pair}
Let $K$ be a CM field with $[K:\Q]=2n$. Its $2n$ embeddings to $\C$ can be grouped into 
$n$ pairs, $(\rho_i, \bar{\rho}_i)$ for $i=1,\cdots, n$, so that $\bar{\rho}_i(x) = (\rho_i(x))^{\rm cc}$, 
where the superscript ${\rm cc}$ is the complex conjugation operation in $\C$. 
To see this, let $K=\Q(\theta)$ for some primitive element $\theta \in K$. For a minimal polynomial 
$p_\theta(x) \in \Q[x]$ for $\theta$, all the $2n$ roots of $p_\theta(x)=0$ have non-zero imaginary 
parts, and are grouped into $n$ pairs, $(\xi_i,\xi_i^{\rm cc})$ with $i=1,\cdots, n$. 
The embedding $\rho_i: \theta \mapsto \xi_i$ forms a pair with 
$\bar{\rho}_i: \theta \mapsto \xi_i^{\rm cc}$.
\end{props}

\begin{exmpl}
Because the extension degree of a CM field is always even, the simplest CM field is a quadratic 
extension over $\Q$; quartic extensions come next. \\
CM fields $K$ with $[K:\Q]=2$ are always in the form of $K\cong \Q[x]/(x^2+d) \cong \Q(\sqrt{-d})$, where 
$d$ is a positive integer that is not divisible by the square of an integer. Fields defined by $K = \Q[x]/(ax^2+bx+c)$
for $a, b, c\in \Q$ with $4ac-b^2 > 0$ can always be brought into the form of $\Q[x]/(x^2+d)$ by redefining 
$x$. Such fields are called \dfn{quadratic 
imaginary field}s. Two embeddings $\rho_{\pm}$ send $x$ to $\pm i\sqrt{d} \in \C$. For 
quadratic imaginary fields, $K^{\rm nc} \cong K$. 
\end{exmpl}

\begin{exmpl}
\label{exmpl:quartic}
A CM field $K$ with $[K:\Q] = 4$ is always in the form of $K \cong K_0[x]/(x^2-p-q\eta)$, $K_0 = \Q[\eta]/(\eta^2-d)$ for a positive square free integer $d$, and $p, q \in \Q$, satisfying $p \pm q\sqrt{d} < 0$.
The last condition needs to be imposed both for $+$ and $-$, because the condition (iii) would not be 
satisfied if $p+q\sqrt{d} < 0$ but $p-q\sqrt{d} > 0$ (or vice versa). 
\end{exmpl}

CM fields $K$ with $[K:\Q]=4$ are not always Galois over $\Q$. It is Galois (i.e., $K^{\rm nc} \cong K$)
if and only if $(p^2-d q^2) = r^2$ for some ${}^\exists r \in \Q$, or $(p^2-dq^2)=d s^2$ for some
$^\exists s \in \Q$. 
When $q = 0$, in particular, $K$ is 
Galois, $K=\Q(\sqrt{-p}, \sqrt{d})$ and ${\rm Gal}(K/\Q) \cong \Z/(2\Z) \times \Z/(2\Z)$. 
See Ex. 8.4 (2) of \cite{shimura1961complex, shimura2016abelian} for more information.

\begin{exmpl}
Any cyclotomic field $K = \Q(\zeta_m)$ is a CM field. $K_0=\Q(\zeta_m + \zeta_m^{-1})$. 
\end{exmpl}

\begin{anythng}
When $K$ is a CM field, its normal closure $K^{\rm nc}$ is also a CM field (Prop. 5.12, \cite{ShimuraIntro}).
\end{anythng}

%%%%%%%%%%%%%%%%%%%%%%%%%%%%%%%%%%%%%%%%%%%%%%%%%%%%%%%%%%%%%%%%
\section{Hodge Structure with Complex Multiplication}
\label{sec:Hodge}
%%%%%%%%%%%%%%%%%%%%%%%%%%%%%%%%%%%%%%%%%%%%%%%%%%%%%%%%%%%%%%%%

%%%%%%%%%%%%%%%%%%%%%%%%%%%%%%%%%%%%%%%%
\subsection{CM-type Varieties and Hodge structures} \label{ss:cmperiod}
%%%%%%%%%%%%%%%%%%%%%%%%%%%%%%%%%%%%%%%%

A K\"{a}hler manifold is specified by a triplet of data $(M, h, J)$, where $M$ is a manifold, 
$J$ the K\"{a}hler form on $M$ and $h$ the complex structure. The complex structure $h$ 
is encoded by specifying a decomposition 
\begin{align}
 H^k(M; \Q) \otimes_\Q \C \cong \oplus_{p+q=k} H^{p,q}
\end{align}
into $h^{p,q}$-dimensional vector spaces---$(p,q)$-Hodge components---over $\C$.

Hodge structure is a notion that extracts the property of complex structure above
in the language of linear algebra. 
A rational Hodge structure\footnote{In this article, we only need to deal with pure 
Hodge structures, since we exclusively deal with compact and smooth geometry for 
compactifications.} $h$ on a vector space $V_\Q$ over $\Q$ with {\bf weight} $k$ is 
a decomposition  
\begin{align}
V_\C:=V_\Q \otimes_\Q \C = \mathop{\bigoplus_{p+q=k}}_{p\geq 0,\; q\geq 0} V^{p,q},
\qquad \overline{V^{p,q}}=V^{q,p}.
\end{align}
Information equivalent to the decomposition is also provided by specifying a representation
\begin{align}
 \tilde{h}: \C^\times \cong {\rm Res}_{\C/\R} (\mathbb{G}_m) \rightarrow {\rm GL}(V_\R)
 \label{eq:htilde}
\end{align}
on $V_\R = V_\Q \otimes \R$; for $(a+ib) \in \C^\times$, $\tilde{h}(a+ib)|_{V^{p,q}}$ multiplies 
$(a+ib)^p(a-ib)^q$. Once such a representation is given, then $V^{p,q}$ can be extracted 
as the eigenspace of the representation $(p,q)$, and the weight is read out through 
$\tilde{h}(a+i0) = a^{p+q}=a^k$.

\begin{defn}
A rational Hodge structure on $V_\Q$ is said to be \dfn{simple}, if there is no rational Hodge substructure.
\dfn{A rational Hodge substructure exists} in a rational Hodge structure on $V_\Q$, when there is a 
vector subspace $W_\Q \subset V_\Q$ over $\Q$ so that $W^{p,q} := (W_\Q \otimes \C) \cap V^{p,q}$ satisfy 
$(W_\Q \otimes \C) \cong \oplus_{p+q = k} W^{p,q}$. An existence of such a $\Q$-vector subspace is highly 
non-trivial; the N\'eron-Severi lattice of the second cohomology of an algebraic K3 surface is an example of such
a subspace, but it does not exist for generic (non-algebraic) complex analytic K3 surfaces.
\end{defn}

\begin{props}
Suppose that there is a rational Hodge substructure on $W_\Q \subset V_\Q$.
Let us then take a vector subspace $U_\Q$ so that $W_\Q \oplus U_\Q \cong V_\Q$. With the definition 
$U^{p,q}:=(U_\Q \otimes \C) \cap V^{p,q}$, the Hodge decomposition on $V_\Q \otimes \C$ can be 
split into those of $W_\Q \otimes \C$ and $U_\Q \otimes \C$; 
\begin{align}
V^{p,q} \cong W^{p,q} \oplus U^{p,q}, \quad (W_\Q \otimes \C) \cong \oplus_{p,q=k} W^{p,q}, \quad 
(U_\Q \otimes \C) \cong \oplus_{p+q=k} U^{p,q}.
\end{align}
\end{props}

\begin{defn}
When a rational Hodge structure on $V_\Q$ of weight $k$ is not simple, its rational 
Hodge substructure on a vector space $W_\Q$ is said to be \dfn{level}-$\ell$, if 
${\rm max}(|p-q|) = \ell$ on $W_\Q\otimes \C \cong \oplus_{p+q=k} W^{p,q}$.
\end{defn}

\begin{anythng}
A compact K\"{a}hler manifold $M$ (with its complex structure specified) introduces a rational 
Hodge structure of weight $k$ on the cohomology group $H^k(M;\Q)$, but, in general, not all 
the mathematically possible rational Hodge structures on the vector space $H^k(M;\Q)$ are realized 
as complex structures of a family of K\"{a}hler manifolds with a given topology. While all the rational 
Hodge structures are realized by complex structures in the case of elliptic curves and K3 surfaces, 
that is not the case for Calabi--Yau threefolds. 
It is still useful to extract the key properties of complex structures of geometries and distil in 
the form of rational Hodge structure, because all the properties of rational Hodge structure are 
satisfied by complex structures of geometries. 

At a generic point in the complex structure moduli space of a family of Calabi--Yau threefolds 
(with $b_1(M)=b_5(M)=0$), we expect that the rational Hodge structure on $H^3(M;\C)$ is simple. 
This does not rule out a possibility that the rational Hodge structure on $H^3(M;\Q)$ stops 
being simple at a sublocus in the moduli space; we encounter such examples in the main text.
\end{anythng}

%%%%%%%%%%%%%%%%%%%%%%%%%%%%%%%%%%%%%%%%%%%%%%%%%%%%%%%%%%%%%%
\subsubsection*{Complex multiplication on an elliptic curve}
%%%%%%%%%%%%%%%%%%%%%%%%%%%%%%%%%%%%%%%%%%%%%%%%%%%%%%%%%%%%%%

\begin{anythng}
The condition for the existence of complex multiplication for an elliptic curve 
$E_\tau$, defined by eq. (\ref{eq:quadratic-eq-tau}), can be translated into 
the language of Hodge structure on the cohomology group $H^1(E;\Q)$. The algebra 
of endomorphisms of the simple rational Hodge structure 
\begin{align}
 K := {\rm End}_{\rm Hdg}(H^1(E;\Q)) := \left\{ \varphi \in {\rm End}_\Q(H^1(E;\Q)) \; | \; 
   \varphi(H^{1,0}) \subset H^{1,0}, \; \varphi(H^{0,1}) \subset H^{0,1} \right\}
\label{eq:endmrph-field-def-ell}
\end{align}
is known to be a field for any complex structure of $E_\tau$, and moreover, 
isomorphic either to $\Q$ or to a quadratic imaginary field (that is, $K$ is 
a CM field with $[K:\Q] = \dim_\Q (H^1(E;\Q))$ ---(**). The condition for 
existence of complex multiplication is known to be equivalent to (**); 
$K \cong \Q(\sqrt{b^2-4ac})$ with $b^2-4ac < 0$ then. 
See \cite{moore1998attractors, moore1998arithmetic, gukovvafarcft, 
moore2004houches}, or math literatures for more information.
We say that $\tau \in {\cal M}_{\rm cpx}^{\rm ell.}$ is a \dfn{CM point} when the 
corresponding elliptic curve $E_\tau$ has complex multiplications (is of CM-type).
\end{anythng}

\begin{anythng}
\label{statmnt:distr-CM-ell}
The CM points ${\cal M}_{\rm CM}^{\rm ell.} \subset {\cal M}_{\rm cpx}^{\rm ell.} = ({\rm SL}(2;\Z) 
\backslash {\cal H}_{g=1})$, where ${\cal H}_{g=1}$ is the upper complex half plane, are classified 
by exploiting symmetry group actions on them as follows. While all the solutions to non-trivial quadratic 
polynomials (\ref{eq:quadratic-eq-tau}) give rise to CM points, and hence there are infinitely 
many CM points, action of well-motivated symmetry groups are not transitive on all of them. 
The CM points are classified by their CM fields $K$, first; in the case of elliptic curves, 
$K$ must be one of quadratic imaginary fields $\Q(\sqrt{-d})$ with a positive square-free 
integer $d$. Furthermore, all the CM points in ${\rm SL}(2;\Z) \backslash {\cal H}_{g=1}$ 
with a given CM field $K = \Q(\sqrt{-d})$ forms a single zero-dimensional Shimura variety 
${\rm Sh}( (\Q(\sqrt{-d}))^\times, \tilde{h})$, where $\tilde{h}$ is a homomorphism 
(\ref{eq:htilde}) corresponding to a rational Hodge structure on $H^1(E;\Q)$ with complex 
multiplication in\footnote{It does not matter which $\tilde{h}$ 
we use to specify ${\rm Sh}(K^\times, \tilde{h})$ in the case of elliptic curves
so long as $\tilde{h}$ has this property. } 
$K=\Q(\sqrt{-d})$; conversely, the Shimura variety ${\rm Sh}(K^\times, \tilde{h})$ of 
any quadratic imaginary field $K = \Q(\sqrt{-d})$ has a non-empty share 
in ${\cal M}_{\rm CM}^{\rm ell.}$. Each one of those Shimura varieties consists of infinitely 
many points, because there are infinitely many quadratic polynomial 
equations (\ref{eq:endmrph-field-def-ell}) whose discriminant is $(-d)$ times the square 
of an integer.\footnote{Since there is a supersymmetric upper bound in the three-brane 
charge contribution from three-form fluxes, not all those infinitely many CM points are 
relevant in practical physics questions \cite{DeWolfe}. In this article, we just simply 
ignore this subtlety by tensoring $\Q$.} 

The Galois group ${\rm Gal}(\overline{\Q}/\Q(\sqrt{-d}))$ acts on the Shimura variety 
${\rm Sh}(\Q(\sqrt{-d})^\times, \tilde{h})$; the action is not transitive. The group 
${\rm SL}(2;\Q) \subset {\rm GL}(2;\Q)$ acts on the CM points ${\cal M}_{\rm CM}^{\rm ell.}$ 
as $\Q$-coefficient rational transformations of $\tau$, but it also acts on individual 
Shimura varieties, not in between Shimura varieties ${\rm Sh}(K^\times, \tilde{h})$ of
different quadratic imaginary fields $K$.
\end{anythng}

%%%%%%%%%%%%%%%%%%%%%%%%%%%%%%%%%%%%%%%%%%%%%%%%%%
\subsubsection*{Complex multiplication on a simple Abelian variety}
%%%%%%%%%%%%%%%%%%%%%%%%%%%%%%%%%%%%%%%%%%%%%%%%%%%

\begin{defn}
An Abelian variety $A$ is said to be \dfn{simple} when it has no proper Abelian subvarieties.
\end{defn}

\begin{anythng}
\label{statmnt:Abelian-smpl-CM}
Let $A$ be a simple Abelian variety\footnote{We only consider Abelian varieties with a principal 
polarization in this article.} of $\dim_\C A = g$. Its complex structure $\tau^{ij} \in {\cal H}_g$
determines a weight-1 rational Hodge structure on $H^1(A;\Q)$; here, $\tau^{ij}$ is a $\C$-valued 
$g \times g$ symmetric matrix whose imaginary part is positive definite, and ${\cal H}_g$ the 
set of all such $\tau^{ij}$'s. The rational Hodge structure on $H^1(A;\Q)$ of an Abelian variety 
is known to be simple if and only if $A$ is simple.\footnote{The authors found 1.11.3, 1.11.4 
and 1.12.1 of \cite{gordon1997matters} informative.}
The algebra of endomorphisms of the simple rational Hodge structure,
\begin{align}
 L := {\rm End}_{\rm Hdg}\left( H^1(A;\Q) \right) :=
 \left\{ \varphi \in {\rm End}_\Q\left( H^1(A;\Q) \right) \; | \; \varphi(H^{p,q}) \subset H^{p,q} \right\}
 \label{eq:def-L-Abelian-simple}
\end{align}
is known to be a division algebra,\footnote{Any non-zero element of $L$ has an inverse element in $L$, 
but the multiplication law is not necessarily commutative. This is the definition of a division algebra.}
and the center of $L$ is denoted by $K$.
It is known that $K$ is always a field. Furthermore, there are three cases (e.g., \cite{shimura1963analytic}); 
Type I: $K$ is a totally real field, and $L = K$, Type II/III:\footnote{This second case can be split further 
into two different types, and it is customary in math literatures to think of $L$ in four different types. We are not 
concerned about such a detailed classification in this article, however.} $K$ is a totally real field, 
and $[L:K] = 4$, and Type IV: $K$ is a CM field, $[L:K] = q^2$ for some integer $q$, and 
$q^2 [K:\Q]\mid 2g$. 
A simple Abelian variety $A$ is said to \dfn{have sufficiently many CM(complex multiplication)s}
or be \dfn{of CM-type} when the algebra $L$ over $\Q$ falls into the Type IV, and moreover, $[K:\Q] = 2g$. Obviously 
this definition is a generalization of the condition (**) for elliptic curves (1-dimensional 
Abelian varieties). For a simple Abelian variety $A$ with sufficiently many complex multiplications, 
the division algebra $L$ agrees with its center $K$, because $q = 1$. 
See \cite{shimura1961complex, shimura2016abelian, shimura1963analytic} and references therein, or more 
modern literatures, for more information. 
\end{anythng}

\begin{anythng}
\label{statmnt:CMfield-repr-cohomology}
Now, the $2g$-dimensional vector space $H^1(A;\Q)$ over $\Q$ can be regarded as a 1-dimensional 
vector space over $K$, where the scalar multiplication of $\varphi \in K$ is to apply the endomorphism 
$\varphi$ on $H^1(A;\Q)$. By choosing any non-zero element $e \in H^1(A;\Q)$, one finds an isomorphism 
$K \cong H^1(A;\Q)$ as a vector space over $K$. A $\Q$-basis $\{ e_{i=1,\cdots, 2g} \}$ 
of $H^1(A;\Q)$ can readily be regarded as a $\Q$-basis $\{ \omega_{i=1,\cdots, 2g}\}$ of $K$ and vice versa.
With this identification, $[A(\varphi)]_{ji}$ in (\ref{eq:multiply-reltn}) is regarded as the 
$\Q$-valued matrix representation (defining representation) of $\varphi \in K$ on $H^1(A;\Q)$. 
Equation (\ref{eq:multiply-reltn-embdd}) implies that $\rho_a(\varphi) \in \C$ with $a=1,\cdots, 2g$ 
are the eigenvalues of the matrix $[A(\varphi)]_{ji}$, and the $2g$ column vectors of the matrix 
$[\rho_a(\omega_i)]^{-1}$ labeled by $a$ are the corresponding eigenvectors of $\rho_a(\varphi)$. 
One and the same eigenspace decomposition of $H^1(A;\Q) \otimes \C$ is shared by all the endomorphisms 
$\varphi = A(\varphi) \in K$.
\end{anythng}

\begin{anythng}
\label{statmnt:def-CMtype-AbelianVar}
Among the $2g$ eigenspaces, $g$ eigenspaces should correspond to $H^{0,1} \subset H^1(A;\Q) \otimes \C$ and the remaining $g$ of them 
to $H^{1,0} \subset H^1(A;\Q) \otimes \C$. The $g$ embeddings corresponding to the former set of eigenspaces 
are grouped into a set $\Phi := \{ \rho_{a=1,\cdots, g} \}$, and the remaining $g$ embeddings into 
$\overline{\Phi} := \left\{ \bar{\rho}_{a=1,\cdots, g} \right\}$. The set of information 
$(K, \Phi)$ of a simple Abelian variety $A$ with sufficiently many complex multiplications 
is called the \dfn{CM type} of $A$. More generally, for a CM field $K$ and a set of its embeddings into $\Qbar$,
$\Phi=\b{\rho_{a=1,\dots,[K:\Q]/2}}$, the pair $(K,\Phi)$ is called a \dfn{CM type} when $\Phi$ contains
no pair of embeddings which are complex conjugate to each other.

For a CM type $(K,\Phi)$, a subfield 
\begin{align}
 K^r := \Q \left( \left\{ \sum_{\rho \in \Phi} \rho(x) \; | \; x \in K \right\} \right)
\end{align}
of $\overline{\Q}$ is called the \dfn{reflex field of} $(K,\Phi)$. By definition, 
it is a subfield of $K^{\rm nc}$. In the case of simple Abelian variety $A$ with sufficiently 
many CMs, $K^{\rm nc}$, $K$ and $K^r$ are not necessarily isomorphic to one another, 
when $\dim_\C A = g > 1$. 
\end{anythng}

%
%
%
%
%
%
%

%%%%%%%%%%%%%%%%%%%%%%%%%%%%%%%%%%%%%%%%%%%%%%%%%%
\subsubsection*{Complex multiplication on an Abelian variety not necessarily simple}
%%%%%%%%%%%%%%%%%%%%%%%%%%%%%%%%%%%%%%%%%%%%%%%%%%%

\begin{defn}
Let $A$ and $B$ be abelian varieties. A homomorphism $A \to B$ is called an \dfn{isogeny}
if it is surjective, and has finite (zero-dimensional) kernel. The existence of an isogeny
$A\to B$ is an equivalence relation between $A$ and $B$. When such an isogeny exists,
$A$ and $B$ are said to be \dfn{isogenous} to each other.
\end{defn}

\begin{anythng}
For any Abelian variety $A$, there is an isogeny with a product of simple Abelian varieties
of the form $(B_1 \times \cdots \times B_1) \times (B_2 \times \cdots \times B_2) \times \cdots 
= B_1^{h_1} \times B_2^{h_2} \times \cdots$, where $B_1,B_2,\dots$ are simple Abelian varieties
and are not isogenous to each other. 
The Abelian variety $A$ is simple if and only if just one simple Abelian variety $B_1$ is found 
in this product, and moreover, $h_1=1$. 

The algebra $L_A$ of endomorphisms preserving the Hodge structure of an Abelian variety $A$ 
depends only on its isogeny class. $L_A$ is a division algebra if and only if $A$ is simple. 
When $A$ is not simple, the algebra has the form of 
\begin{align}
 L_A \cong M_{h_1}(D_1) \times M_{h_2}(D_2) \times \cdots, \qquad \qquad 
  D_i = {\rm End}_{\rm Hdg}(B_i),
\end{align}
where $M_h(D)$ is the algebra of $D$-valued $h \times h$ matrices. Let $K_i$ be the 
center of the division algebra $D_i$; the center $K_i$ is a field (as in \ref{statmnt:Abelian-smpl-CM}), and it is also
the center of $M_{h_i}(D_i)$. It is further known \cite{Reiner,bilucomplex}
that there always exists 
a maximal subfield $F_i$ of $M_{h_i}(D_i)$ containing $K_i$, where $[F_i: K_i] = h_i q_i$.
\end{anythng}

\begin{anythng}
\label{statmnt:def-CM-Abelian}
An Abelian variety that is not simple is said to {\bf have sufficiently many CMs}  
or {\bf be of CM-type},  
if and only if all the simple Abelian varieties $B_i$ have sufficiently many CMs. 
In this case, all the fields $F_i$ are CM fields, and $[F_i:\Q] = 2 h_i \dim_\Q B_i$.
\end{anythng}

\begin{anythng}
\label{statmnt:def-isotypic}
An Abelian variety $A$ is said to be {\bf isotypic}, if it is isogenous to a product 
$B^h$ of a simple Abelian variety $B$. When an isotypic Abelian variety $A$ is of CM type, 
the set $(F, \Phi_F)$, where $F$ is its CM field and $\Phi_F$ the half of the embeddings 
of $F$ to $\C$ corresponding to $H^{1,0}(A)$, is said to be its {\bf CM type}. The reflex 
field of $(F, \Phi_F)$ is defined as in \ref{statmnt:def-CMtype-AbelianVar}.
\end{anythng}

\begin{anythng}
\label{statmnt:reflex-1}
A CM type $(K, \Phi)$ is said to be \dfn{primitive}, if there is no non-trivial subfield 
$K'$ of $K$ and its CM type $(K', \Phi')$ from which $\Phi$ is induced. ($\Phi$ is induced 
by inspecting whether an embedding of $K$ falls into $\Phi'$ or $\overline{\Phi}'$ when 
restricted upon $K'$). For a CM field $F$ associated with an Abelian variety $(B)^{h}$ 
made up of a simple Abelian variety $B$ with sufficiently many CMs, for example, 
a CM type $(F, \Phi_F)$ is induced from the CM type $(K,\Phi)$ of $B$; $(F,\Phi_F)$ and $(K, \Phi)$ 
play the role of $(K, \Phi)$ and $(K',\Phi')$, respectively, in the definition. \\
This characterization of the primitiveness/non-primitiveness of a CM type in terms 
Abelian varieties can also be used as an alternative definition. This is possible because
for a given CM type $(F, \Phi_F)$, one can always find an Abelian variety $A$ 
whose CM type is $(F, \Phi_F)$, as we will state in \ref{statmnt:Shimura4Abelian}.
When $(K,\Phi)$ is not primitive, the presence of the CM type $(K',\Phi')$ with 
$h:= [K:K']>1$ infers that Abelian varieties for $(K,\Phi)$ are not simple, but 
isogenous to $B^h$ for some Abelian variety $B$ for $(K',\Phi')$. Conversely, when 
the Abelian variety $A$ for a CM type $(K,\Phi)$ is not primitive, an Abelian subvariety 
$B$ of $A$ has a CM type $(K',\Phi')$ from which $(K,\Phi)$ is induced. 
\end{anythng}

\begin{anythng}
When a CM type $(K, \Phi)$ is given, then there is a notion of unique primitive CM subtype 
$(K',\Phi')$, from which $(K, \Phi)$ is induced \cite{milne2006complex}. 
$(K', \Phi')$ is constructed by first thinking of a CM type $(K^{\rm nc}, \Phi^{\rm nc})$, 
where $\Phi^{\rm nc}$ is induced from $\Phi$, and then determine $(K', \Phi')$ as the unique 
primitive CM subtype of $(K^{\rm nc}, \Phi^{\rm nc})$, which is also the unique primitive 
CM subtype of $(K, \Phi)$. $(K, \Phi)$ is primitive, if and only if $K=K'$.
\end{anythng}

\begin{anythng}
For a CM type $(K, \Phi)$, one can think of a CM type $(K^{\rm nc}, (\Phi^{\rm nc})^{-1})$, and then 
its unique primitive CM subtype, which is denoted by $(K^r, \Phi^r)$. The CM field characterized 
in this way is denoted by $K^r$, the same as the reflex field of $(K, \Phi)$, because they are 
actually the same. The CM type $(K^r, \Phi^r)$ is called the \dfn{reflex} of the CM type $(K, \Phi)$.
\end{anythng}

\begin{anythng}
\label{statmnt:Krr-Kprime}
The reflex of the reflex $(K^r, \Phi^r)$ of a CM type $(K,\Phi)$ is denoted by $(K^{rr}, \Phi^{rr})$; 
it is known that $K^{rr} \subset K$, and $\Phi$ is induced from $\Phi^{rr}$. 
Since $(K^{rr},\Phi^{rr})$ is always primitive, by construction, it follows that
$(K',\Phi') = (K^{rr}, \Phi^{rr})$.  
% A CM type $(K, \Phi)$ is primitive if and only if $(K^{rr}, \Phi^{rr}) = (K, \Phi)$. 
\end{anythng}

\begin{anythng} 
\label{statmnt:notseenelsewhere-1}
The reflex $(K^r, \Phi^r)$ of a CM type $(K, \Phi)$ agrees with that 
of $(K',\Phi')$, where $(K',\Phi')$ is the unique primitive CM subtype of $(K, \Phi)$ (\S 20.1, \cite{shimura2016abelian}). 

Let $A$ be an Abelian variety that is isogenous to $B^h$, where $B$ is a simple Abelian variety, 
and $(F, \Phi_F)$ [resp. $(K, \Phi)$] be the CM type of $A$ [resp. $B$]. 
The reflex $(K^r, \Phi^r)$ of $(F, \Phi_F)$ therefore depends only on $(K, \Phi)$ of the simple 
Abelian variety. 
\end{anythng}

\begin{anythng}
The reflex CM type $(K^r, \Phi^r)$ is not always the same as $(K, \Phi)$. 
If $K^{\rm nc}/\Q$ is an Abelian extension, then $K'=K^r$. 
If $(K,\Phi)$ is primitive, then $K'=K$. So the combination of those two conditions 
is a sufficient condition for $K=K^r$ (Ex. 8.4. (1), \cite{shimura1961complex, shimura2016abelian}). 
All the quadratic imaginary fields are covered by this sufficient condition 
for $(K, \Phi) = (K^r, \Phi^r)$. 
\end{anythng}

\begin{defn}
\label{statmnt:reflex-x}
For a CM type $(K, \Phi)$, the {\bf type norm} is the group homomorphism 
$N_\Phi: K^\times \ni x \mapsto \prod_{\rho \in \Phi}(\rho(x)) \in (K^r)^\times$. 
The {\bf reflex norm} is the group homomorphism 
$N_{\Phi^r}: (K^r)^\times \ni y \mapsto \prod_{\rho' \in \Phi^r} \rho'(y) \in (K^{rr})^\times \subset K^\times$. 
\end{defn}

%
%
%
%
%
%
%

%%%%%%%%%%%%%%%%%%%%%%%%%%%%%%%%%%%%%%%%%%%%%%%%%%%%%%%%%%%%%
\subsubsection*{Classification of CM points in ${\cal H}_g$}
%%%%%%%%%%%%%%%%%%%%%%%%%%%%%%%%%%%%%%%%%%%%%%%%%%%%%%%%%%%%%

\begin{anythng}
\label{statmnt:Shimura4Abelian}
Here is a statement on the classification of $g$-dimensional Abelian varieties with 
sufficiently many CMs; this is the generalization of the statement \ref{statmnt:distr-CM-ell} 
for elliptic curves. 
First, any CM type $(K, \Phi)$ with $[K:\Q] = 2g$ specified abstractly can be realized as the CM type of an 
Abelian variety. Such an Abelian variety can be constructed by taking a quotient of 
$K \otimes_\Q \R \cong \R^{2g}$ by an ideal of a ring of algebraic integers within $K$, and 
by introducing complex structure on this torus by $\Phi$. There can be many Abelian varieties 
that are not mutually isomorphic for a given CM type $(K,\Phi)$, because there is freedom 
in choosing the ideal \cite{milne2006complex, bilucomplex, shimura1961complex, shimura2016abelian}. 
A zero-dimensional subvariety ${\rm Sh}(N_{\Phi^r}((K^r)^\times),\tilde{h})$ in the moduli space
${\rm Sp}(2g;\Z) \backslash {\cal H}_g = {\cal M}_{\rm cpx}^{A_g}$ consists of the corresponding CM points
of a given CM type $(K, \Phi)$.
Any CM type $(K, \Phi)$ with $[K:\Q] = 2g$ has its non-empty share of CM points in the form of 
${\rm Sh}(N_{\Phi^r}((K^r)^\times), \tilde{h})$ in ${\cal M}_{\rm CM}^{A_g}$.
Unlike in the case of elliptic curves, however, it is not guaranteed whether there is just 
one Shimura variety associated with a given CM field $K^r$ in ${\cal M}_{\rm CM}^{A_g}$, or maybe 
more than one with the same $K^r$.

Note also that this statement is valid whether the CM type $(K,\Phi)$ is primitive or not; 
the Mumford--Tate group (\ref{eq:MT-reflex-rltn}) is not necessarily of $2g$-dimensions. 

The Galois group ${\rm Gal}(\overline{\Q}/K^r)$ acts on individual Shimura 
varieties ${\rm Sh}(N_{\Phi^r}((K^r)^\times),\tilde{h})$; the action is not 
transitive. The action of ${\rm Sp}(2g;\Q)$ also preserves the CM type 
$(K, \Phi)$. The action of ${\rm Sp}(2g;\Q)$ illustrates nicely that 
the CM points arise as a family of infinitely many.
See \cite{edixhoven2003subvarieties, clozel2005equidistribution, zhang2005equidistribution, ullmo2007manin, ullmo2012galois}
for developments on the distribution of CM points in the moduli space. 
\end{anythng}

%%%%%%%%%%%%%%%%%%%%%%%%%%%%%%%%%%%%%%%%%%%%%%%%%%%%%%%%%%%%%
\subsubsection*{CM-type K3 surface} 
%%%%%%%%%%%%%%%%%%%%%%%%%%%%%%%%%%%%%%%%%%%%%%%%%%%%%%%%%%%%%

Let $T$ be the transcendental lattice of a K3 surface $S$. $T$ is a rank-$(22-\rho)$ lattice, 
where $\rho$ is the Picard number of $S$. 
$T\otimes \Q$ admits a weight-2 simple rational Hodge substructure of $H^2(S;\Q)$. 
The algebra of endomorphisms of the rational Hodge structure 
\begin{align}
 K := {\rm End}_{\rm Hdg} \left( T \otimes \Q \right) :=
  \left\{ \varphi \in {\rm End}_\Q \left( T \otimes \Q \right) \; | \; \varphi(H^{p,q}) \subset H^{p,q} \right\}
 \label{eq:def-K-K3}
\end{align}
is a field. It is known that $K$ is either totally real or a CM field, and that 
$\dim_\Q (T \otimes \Q) = (22-\rho)$ is divisible by $[K:\Q]$. A K3 surface $S$ is 
said to be a \dfn{CM-type K3 surface} if the field $K$ is a CM field with 
$[K:\Q] = \dim_\Q (T \otimes \Q)$. Once again, this definition is a generalization 
of (**) for elliptic curves. Since there is no operation of multiplying a complex number in certain 
way as an action from $S$ to $S$ in this definition,\footnote{Kuga--Satake construction assigns an isogeny 
class of $2^{r'-2}$-dimensional Abelian varieties $[KS(T_S)]$ to a rank-$r'$ transcendental lattice 
with a rational Hodge structure of K3 type. $S$ is a CM-type K3 surface if and only if 
the Abelian varieties $[KS(T_S)]$ have sufficiently many CMs \cite{huybrechts2016lectures, tretkoff2015k3}.
Complex multiplications are still realized geometrically on the Abelian varieties $[KS(T_S)]$, though not 
on the K3 surface $S$ itself.} 
we no longer use ``complex multiplication'' but just say ``CM.'' 
Due to the definition of a CM field, $(22-\rho)$ is an even number for a CM-type K3 surface.
See (\S 3, \cite{huybrechts2016lectures}) and \cite{zarhin1983hodge, van2006real}
for more information. 

Precisely the same statement as in \ref{statmnt:CMfield-repr-cohomology} holds true for the CM field $K$ 
of a CM-type K3 surface $S$, after replacing $H^1(A;\Q)$ by $T \otimes \Q$ and its dimension $2g$
by $22-\rho$. A statement for a CM-type K3 surface corresponding to \ref{statmnt:def-CMtype-AbelianVar} 
is the following: among the $22-\rho$ embeddings of $K$ into $\C$, two embeddings correspond to 
the eigenspaces $H^{2,0}$ and $H^{0,2}$ of all $A(\varphi) = \varphi \in K$; those embeddings are denoted by 
$\epsilon$ and $\bar{\epsilon}$, respectively; all the remaining $(20-\rho)$ eigenspaces combined 
correspond to the $H^{1,1}$ component of $T \otimes \C$.  

It is known that any CM field $K$ with $[K:\Q]=2n$, $2 \leq 2n \leq 20$, can be realized 
as the field $K$ of a K3 surface $S$ defined by (\ref{eq:def-K-K3}) \cite{pjateckii1975arithmetic,taelman2015K3-Lfcn}; 
the transcendental lattice $T$ of $S$ then has a rank $2n$. The proof of this statement in \cite{pjateckii1975arithmetic}
(which is for $2n \leq 16$) constructs, for a CM field $K$ and one of its embeddings $\epsilon$, 
$\Q$-valued intersection forms on $K$ so that $\epsilon$ is associated with the Hodge $(2,0)$ component; 
a sublattice $T$ needs to be extracted from the vector space $K$
so that the intersection form becomes even and integral. 
The way the intersection forms are constructed in \cite{pjateckii1975arithmetic} suggests strongly 
that there are infinitely many inequivalent rank-$2n$ lattices $T_0$ and its period domain $D(T_0)$ 
where a given pair $(K, \epsilon)$ can be realized. Once a CM point with $(K, \epsilon)$ is found 
in a period domain $D(T_0)$, however, there must be infinitely many CM points with the same $(K, \epsilon)$ 
within $D(T_0)$ of the same lattice $T_0$ (if $2n>2$). This is because the group 
${\rm Isom}(T_0 \otimes \Q)$ takes one CM point in $D(T_0)$ to 
elsewhere within the same $D(T_0)$ without changing the pair $(K, \epsilon)$. 

One will also be interested in finding a list of pairs $(K, \epsilon)$ of CM points that can be realized 
in the period domain of an even lattice $T_0$ with signature $(2,2n-2)$. One just has to invert 
the $T_0 \Rightarrow (K, \epsilon)$ list referred to above, although it is not a simple task to do so 
in practice. At least it is clear that not all possible pairs $(K, \epsilon)$ with 
$[K:\Q] = 2n$ are admitted in a given lattice $T_0$ with signature $(2,2n-2)$. 
Take the $2n=2$ case (i.e., $\rho= 20$), as an obvious example. A rank-2 transcendental 
lattice $T_0$ admits just one CM point, where $K$ is the quadratic imaginary field 
uniquely determined by the transcendental lattice. 
Slightly more non-trivial examples are the cases of $T_0 = U[2] \oplus U[2]$ (when 
$S = {\rm Km}(E \times F)$ Kummer surface associated with mutually non-isogenous 
elliptic curves $E$ and $F$) and $T_0 = U \oplus U$; a CM field with $[K:\Q] = 4$ is 
realized on these two $D(T_0)$'s only when the field $K$ is constructed by setting 
$q=0$ in Example \ref{exmpl:quartic}. To the knowledge of the authors, it is not 
guaranteed whether a CM point is found in the non-Noether--Lefschetz locus\footnote{
\label{fn:NL} A {\bf Noether--Lefschetz locus} of $D(T_0)$ is a subspace of $D(T_0)$ 
where $T \subsetneq T_0$.} in the period domain $D(T_0)$ of a given lattice $T_0$; when 
the rank of $T_0$ is odd, for example, obviously there cannot be a CM point in the 
non-Noether--Lefschetz locus of $D(T_0)$. Once a CM point of a pair $(K, \epsilon)$ is 
found in $D(T_0)$ with $[K:\Q] = \dim_\Q T_0$, then this CM point is not in a 
Noether--Lefschetz locus of $D(T_0)$, because the fact that $K$ is a field 
indicates that the rational Hodge structure on $T_0$ at the CM point is simple. 

%%%%%%%%%%%%%%%%%%%%%%%%%%%%%%%%%%%%%%%%%%%%%%%%%%%%%%%%%%%%%
\subsubsection*{CM-type Calabi--Yau threefold}
%%%%%%%%%%%%%%%%%%%%%%%%%%%%%%%%%%%%%%%%%%%%%%%%%%%%%%%%%%%%%

When the weight-3 rational Hodge structure on $H^3(M;\Q)$ of a Calabi--Yau threefold $M$ 
is simple, the algebra
\begin{align}
 K := {\rm End}_{\rm Hdg}\left( H^3(M;\Q) \right) := \left\{ \varphi \in 
 {\rm End}_\Q(H^3(M;\Q)) \; | \; \varphi ( H^{p,q} ) \subset H^{p,q} \right\}
 \label{eq:def-K-CY3}
\end{align}
is a field. The Calabi--Yau threefold $M$ (and also the rational Hodge structure) is said to be 
\dfn{of CM-type} if $K$ is a CM field with $[K:\Q] = \dim_\Q (H^3(M;\Q))$, which is once again 
a generalization of the condition (**) of elliptic curves. Precisely the same statement as in 
\ref{statmnt:CMfield-repr-cohomology} holds true for the CM field $K$, by replacing $H^1(A;\Q)$
with $H^3(M;\Q)$ and $2g$ with $\dim_\Q(H^3(M;\Q))$, respectively. Among the $\dim_\Q(H^3(M;\Q))$ 
embeddings of $K$ into $\C$, there are two special ones, denoted by $\epsilon$ and $\bar{\epsilon}$, 
whose corresponding eigenspaces are the $H^{3,0}$ and $H^{0,3}$ Hodge components, respectively. 
All the other embeddings correspond to the eigenspaces that fit within either $H^{2,1}$ or $H^{1,2}$ 
Hodge components. 

%%%%%%%%%%%%%%%%%%%%%%%%%%%%%%%%%%%%%%%%%%%%%%%%%%%%%%%%%%%%%
\subsubsection*{CM-type Rational Hodge Structure}
%{\bf CM-type Rational Hodge Structure}
%%%%%%%%%%%%%%%%%%%%%%%%%%%%%%%%%%%%%%%%%%%%%%%%%%%%%%%%%%%%%

When a vector space $V_\Q$ over $\Q$ has a rational Hodge structure that is simple, 
then the algebra of endomorphisms of the simple rational Hodge structure 
\begin{align}
 L := {\rm End}_{\rm Hdg}(V_\Q) := \left\{ \varphi \in {\rm End}_\Q (V_\Q) \; | \; 
   \varphi (V^{p,q}) \subset V^{p,q} \right\}
  \label{eq:def-L-Hdge-simple}
\end{align}
is always a division algebra. When $L$ contains a number field $K$ such that $[K : \Q] = \dim_\Q V_\Q$,
we say that the simple rational Hodge structure is \dfn{of CM-type}. 

When a rational Hodge structure on a vector space $V_\Q$ can be decomposed into 
multiple rational Hodge substructures that are simple, we say that the rational 
Hodge structure is \dfn{of CM-type}, when all the simple substructures
are of CM-type. When a Calabi--Yau threefold $M$ has a complex structure such that 
the rational Hodge structure on $H^3(M;\Q)$ is not simple, it is said to be \dfn{of CM-type}, 
if and only if the rational Hodge structure on $H^3(M;\Q)$ is of CM-type.
We may define Abelian varieties of CM-type that are not necessarily simple
by requiring that  $H^1(A;\Q)$ is of CM-type, and this definition is
equivalent to the one in \ref{statmnt:def-CM-Abelian}.

In this article, we are frequently forced to refer to the field $K$ as 
the algebra of endomorphisms of a simple Hodge structure of Calabi--Yau type 
(\ref{eq:endmrph-field-def-ell}, \ref{eq:def-K-K3}, \ref{eq:def-K-CY3}), 
the ceter $K$ of the division algebra $L$ for a simple Abelian variety 
in (\ref{eq:def-L-Abelian-simple}), the field $K$ contained in the algebra $L$ 
of a simple rational Hodge structure (\ref{eq:def-L-Hdge-simple}) and the 
maximal subfield $F$ contained in the central simple algebra $M_h(D)$ for 
an isotypic Abelian variety. We just use the phrase \dfn{endomorphism field} 
for all of them in this article (as many literatures also do), since the proper 
expressions such as ``the field of endomorphisms of a simple rational Hodge structure''
is too long. This is not too much abuse of language for Abelian varieties in the 
first place, since the $\Q$-endomorphisms preserving the Hodge structure on 
$H^1(A;\Q)$ can be regarded as some endomorphisms of a group variety $A$ tensored 
with $\Q$.   

%%%%%%%%%%%%%%%%%%%%%%%%%%%%%%%%%%%%%%%%%%%%%%%%%%%%%%%%%%%%%
\subsection{Hodge Components and Embeddings of a CM Field}
\label{ssec:reltn-Hdg-cmp-embd}
%%%%%%%%%%%%%%%%%%%%%%%%%%%%%%%%%%%%%%%%%%%%%%%%%%%%%%%%%%%%%

Suppose that a simple rational Hodge structure is given on a vector space $V_\Q$, 
and that the endomorphism field $K$ is a CM field satisfying $[K:\Q]= \dim_\Q V_\Q =: m$. 
We have seen in \ref{statmnt:CMfield-repr-cohomology} that, for $a=1,\cdots, m$ 
labeling distinct embeddings $\rho_a: K \hookrightarrow \C$, we can choose a vector 
$v_a \in V_\Q \otimes \C$ so that it is an eigenvector of $A(\varphi) = \varphi \in K$ 
with an eigenvalue $\rho_a(\varphi) \in \C$ for any $\varphi \in K$. Moreover, 
when the vector $v_a$ is expressed as $v_a = \sum_{i=1}^m e_i c^i_a$, where $\{e_{i=1, \cdots, m}\}$ 
is a $\Q$-basis of $V_\Q$, and $c^i_a \in \C$, these coefficients $c^i_a$'s can be 
chosen within $K^{\rm nc} \subset \C$, because all the $m \times m$ entries of the matrix 
$[\rho_a(\omega_i)]^{-1}$ take their values in $K^{\rm nc}$. Although the reasoning here 
is applicable\footnote{The corresponding statement is this:
when multiplication of elements of $F$ is seen as action on the vector space $F$ over $\Q$,  
the $\Q$-representation of $F$ over $F_\Q$ can be decomposed into 1-dimensional (and hence 
irreducible) representations when the representation space $F_\Q$ is tensored with
the splitting field $F^{\rm nc}$.} 
to any finite extension field $F$ over $\Q$, yet in the present context, 
it also means that the Hodge decomposition is possible not just in $V_\Q \otimes \C$, 
but even within $V_\Q \otimes K^{\rm nc}$.

In fact, there is a much stronger result. Let $\{x_{p=1,\cdots, m} \}$ be a $\Q$-basis of $K$. 
In the notation above, 
\begin{align}
 A_{ji}(x_p) c_{ia} = c_{ja} \rho_a(x_p).
\end{align}
We use those relations for $j=1$. Now, thinking that those relations are 
\begin{align}
 \left[ A_{1i}(x_p) \right]_{pi} \left[\frac{c_{ia}}{c_{1a}}\right]_i = \left[\rho_a(x_p)\right]_p,
\end{align}
where a $\Q$-valued matrix $[ \cdots ]_{pi}$ is multiplied to a $\C$-valued vector $[ \;\;  ]_i$ 
to be a $\C$-valued vector $[ \; \;  ]_p$ on the right-hand side, we see that the 
$\Q$-valued matrix must be invertible; this is because $x_p$'s in $K$ (and hence $\rho_a(x_p)$'s in $\rho_a(K)$)
should be linearly independent over $\Q$. So, we replace the $\Q$-basis of $K$ $\{x_{p=1,\cdots, m}\}$ 
by the one---denoted by $\{ y_{i=1,\cdots, m}\}$---obtained by multiply the inverse of the $\Q$-valued matrix 
$[ A_{1i}(x_p) ]_{pi}$ on $x_p$'s. In this new basis, there is a relation 
\begin{align}
  \frac{c_{ia}}{c_{1a}} = \rho_a(y_i).
 \label{eq:key-structure}
\end{align}
The coefficients $c_{ia}/c_{1a}$ for the eigenvector $v_a$ all take their values in 
$\rho_a(K) \subset K^{\rm nc}$, first of all, and $c_{ia}/c_{1a} \in \C$ for $a=1,\cdots, m$ 
are the images of the embeddings $\rho_a$ of a common element $y_i \in K$, secondly. 

When the CM field $K$ is Galois, $\rho_a \circ (\rho_b)^{-1} \in {\rm Gal}(K/\Q)$ maps 
the algebraic number $c_{ib}/c_{1b} \in K \subset \overline{\Q}$ to 
$c_{ia}/c_{1a} \in K \subset \overline{\Q}$ for all $i=1,\cdots, [K:\Q]$ simultaneously, 
as in \cite{DeWolfe}. Even when the CM field $K$ is not a Galois extension over $\Q$, 
an isomorphism $\rho_a \circ (\rho_b)^{-1}: \rho_b(K) \rightarrow \rho_a(K)$ extends 
to an isomorphism from $\overline{\Q}$ to itself (Thm. 2.19, \cite{Fujisaki}), which 
can be restricted to an automorphism of a normal extension $K^{\rm nc}$ over $\Q$. 
Thus, $\rho_a \circ (\rho_b)^{-1}: \rho_b(K) \rightarrow \rho_a(K)$ can be realized 
by restricting some elements in ${\rm Gal}(K^{\rm nc}/\Q)$. Therefore, the simultaneous map 
of algebraic numbers $c_{ib}/c_{1b} \in \rho_b(K)$ to $c_{ia}/c_{1a} \in \rho_a(K)$ 
can be regarded as a result of an automorphism in ${\rm Gal}(K^{\rm nc}/\Q)$.

%%%%%%%%%%%%%%%%%%%%%%%%%%%%%%%%%%%%%%%%%%%%%%%%%%%%%%%%%%%%%%%%%%%%%%%%%%%%%
% \begin{figure}[tbp]
% \begin{center}
%   \includegraphics[width=.8\linewidth]{hyperelliptic_2}  
% \caption{\label{fig:xxxx} 
% }
% \end{center}
% \end{figure}
%%%%%%%%%%%%%%%%%%%%%%%%%%%%%%%%%%%%%%%%%%%%%%%%%%%%%%%%%%%%%%%%%%%%%%%%%%%%%%

\bibliographystyle{JHEPKK} %abbrvは普通の番号 / apalikeは名前と年が表示される / alphaはもう少し数学の論文っぽい
\bibliography{Bibliography}

%%
%\begin{thebibliography}{99}
%
%\end{thebibliography}
%
\end{document}